\documentclass[11pt]{article}
\usepackage{cite}
\usepackage{amsmath,amsfonts,amssymb}
\usepackage[small,bf,hang]{caption}
\usepackage{slashed}
\usepackage{mathabx}
\usepackage{latexsym,epsfig}

\def\hybrid{
        \topmargin -20pt
        \oddsidemargin 0pt
        \headheight 0pt \headsep 0pt
        \textwidth 6.25in 
        \textheight 9.5in 
        \marginparwidth .875in
        \parskip 5pt plus 1pt \jot = 1.5ex}

\hybrid

 \csname
@addtoreset\endcsname{equation}{section}

\def\moth{\mathsurround=0pt}
\newdimen\zo \zo=0pt

\def\tick{\leaders\hrule height 0.5ex depth 0pt \hskip 0.5pt}
\def\upboxfill{$\moth \setbox\zo\hbox{\tick}%
  \hskip 3pt\hbox to 0pt{$\tick$\hss}\hrulefill \hbox to 7.5pt{$\tick$\hss}$}

\def\dtick{\leaders\hrule height .34pt depth 0.5ex \hskip 0.5pt}
\def\downboxfill{$\moth \setbox\zo\hbox{\dtick}%
  \hskip 2pt\hbox to 0pt{$\dtick$\hss}\hrulefill \hbox to 2pt{$\dtick$\hss}$}

\def\bec{\begin{center}}
\def\ec{\end{center}}

\def\be{\begin{equation}}
\def\ee{\end{equation}}
\def\bea{\begin{eqnarray}}
\def\eea{\end{eqnarray}}
\def\ba{\begin{array}}
\def\ea{\end{array}}

\begin{document}

\begin{titlepage}
\rightline{}
\rightline{\tt  MIT-CTP/4522}
\rightline{December 2013}
\begin{center}
\vskip .6cm
{\Large \bf {Exceptional Field Theory II:\, E$_{7(7)}$}}\\
\vskip 1.6cm
{\large {Olaf Hohm${}^1$ and Henning Samtleben${}^2$}}
\vskip .6cm
{\it {${}^1$Center for Theoretical Physics}}\\
{\it {Massachusetts Institute of Technology}}\\
{\it {Cambridge, MA 02139, USA}}\\
ohohm@mit.edu
\vskip 0.2cm
{\it {${}^2$Universit\'e de Lyon, Laboratoire de Physique, UMR 5672, CNRS}}\\
{\it {\'Ecole Normale Sup\'erieure de Lyon}}\\
{\it {46, all\'ee d'Italie, F-69364 Lyon cedex 07, France}}\\
henning.samtleben@ens-lyon.fr

\vskip 1.5cm
{\bf Abstract}
\end{center}

\vskip 0.2cm

\noindent
\begin{narrower}
We introduce exceptional field theory for the group E$_{7(7)}$, based on a 
$(4+56)$-dimensional spacetime subject to a covariant section condition. 
The `internal' generalized diffeomorphisms 
of the coordinates in the fundamental representation of E$_{7(7)}$ 
are governed by a  covariant `E-bracket', which is gauged by 56 vector fields.
We construct the complete and unique set of field equations that is gauge invariant under 
generalized diffeomorphisms in the internal and external coordinates.
Among them feature the non-abelian twisted self-duality equations for
the 56 gauge vectors. 
We discuss the explicit solutions of the section condition describing
the embedding of the full, untruncated 11-dimensional and type IIB supergravity, respectively. 
As a new feature compared to the previously constructed E$_{6(6)}$ formulation,
some components among the 56 gauge vectors descend from the 11-dimensional
dual graviton but nevertheless  allow for a consistent coupling by virtue of a covariantly constrained 
compensating 2-form gauge field.

\end{narrower}

\end{titlepage}

\newpage

\tableofcontents


\bigskip
\bigskip
\bigskip

\section{Introduction}
In this paper we present the details of the recently announced `exceptional field theory' 
(EFT)~\cite{Hohm:2013pua} for the group E$_{7(7)}$, 
complementing the E$_{6(6)}$ covariant construction given in \cite{Hohm:2013vpa}.
The approach is a generalization of double field theory (DFT) 
\cite{Siegel:1993th,Hull:2009mi,Hull:2009zb,Hohm:2010jy,Hohm:2010pp,Hohm:2010xe},\footnote{See \cite{Hohm:2013bwa}
for a review and further references.} 
with the goal to render the dynamics of  the complete $D=11$ supergravity \cite{Cremmer:1978km} covariant
under the exceptional groups that are known to appear under
dimensional reduction~\cite{Cremmer:1979up}. 
We refer to the introduction of \cite{Hohm:2013vpa} for a more detailed outline of the general ideas,
previous attempts, and extensive references. Here we will mainly present and discuss 
the novel aspects relevant for the larger group  E$_{7(7)}$.

The E$_{7(7)}$ EFT is based on a generalized $4+56$ dimensional spacetime, with 
the `external' spacetime coordinates $x^{\mu}$ and `internal' coordinates $Y^M$
in the fundamental representation $\bf{56}$ of E$_{7(7)}$, 
with dual derivatives $\partial_M$.\footnote{
Such generalized spacetimes also appear in the proposal of~\cite{West:2003fc}.}
Correspondingly, the field content incorporates an external frame field (`vierbein') $e_{\mu}{}^{a}$ 
and an internal generalized metric ${\cal M}_{MN}$, parametrizing the coset space 
 E$_{7(7)}/$SU$(8)$. Crucially, the theory also requires the presence 
 of generalized gauge connections $A_{\mu}{}^{M}$ and a set of 2-forms $\{B_{\mu\nu\,\alpha}\,,\, B_{\mu\nu\,M}\}$,
  in order to consistently describe the complete degrees of freedom 
 of $D=11$ supergravity (and necessarily including also some of their duals).  
The 2-forms $B_{\mu\nu\,\alpha}$ in the adjoint representation of E$_{7(7)}$ are known from
the dimensionally reduced theory where they show up as the on-shell duals of the four-dimensional scalar fields.
 The significance of the  additional two-forms $B_{\mu\nu\,M}$ in the fundamental representation 
  will become apparent shortly. The presence of these fields that go beyond the field content
  of the dimensionally reduced theory, is required for gauge invariance (under generalized diffeomorphisms) 
  and at the same time crucial in order to reproduce the full dynamics of $D=11$ supergravity.   
  All fields are subject to a covariant section constraint which implies that only a subset of the $56$ 
 internal coordinates is physical. The constraint can be written in terms of the E$_{7(7)}$ generators $(t_\alpha)^{MN}$ 
 in the fundamental representation, and the invariant symplectic form  $\Omega_{MN}$ of E$_{7(7)}\subset {\rm Sp}(56)$,
 as
 \be
 \begin{split}
  (t_\alpha)^{MN}\,\partial_M \partial_N A \ &= \ 0\;, \qquad   (t_\alpha)^{MN}\,\partial_MA\, \partial_N B \ = \ 0 \,,\\
  \Omega^{MN}\,\partial_MA\, \partial_N B \ &= \ 0 \,, 
 \end{split}
 \label{sectioncondition}
 \ee  
for any fields or gauge parameters $A,B$. 

Our main result is the construction of the gauge invariant E$_{7(7)}$ EFT with the field content 
described above 
\bea
\left\{e_\mu{}^a\,,\; {\cal M}_{MN}\,,\; A_\mu{}^M\,,\; B_{\mu\nu\,\alpha}\,,\; B_{\mu\nu\,M} \right\}
\;. 
\label{fieldcontent}
\eea
The 56 gauge fields $A_\mu{}^M$ are subject to the first-order twisted self-duality equations
\bea
{\cal F}_{\mu\nu}{}^M &=& -\frac12\,e\,\varepsilon_{\mu\nu\rho\sigma}\,\Omega^{MN}{\cal M}_{NK}{\cal F}^{\rho\sigma\,K}
\;,
\label{duality56}
\eea
with properly covariantized non-abelian field strengths ${\cal F}_{\mu\nu}{}^M$ that we will introduce below.
In the abelian limit and upon dropping the dependence on all internal coordinates $Y^M$, 
these duality equations are known from the dimensional reduction of $D=11$ supergravity 
to four spacetime dimensions~\cite{Cremmer:1979up}. In that case, they provide a duality
covariant description of the dynamics of the gauge field sector. In particular, 
 after the choice of a symplectic frame, these equations readily encode 
the standard second order field equations for the 28 electric vector fields.
On the other hand, 
the full non-abelian self-duality equations (\ref{duality56}) that we present in this paper reproduce 
the dynamics of the full (untruncated) eleven-dimensional supergravity for these fields.

In addition to (\ref{duality56}), the dynamics of the remaining fields is described by second order field equations, that are
most conveniently derived from an action: 
\bea
\label{finalaction}
\begin{split}
 S_{\rm EFT} \ = \  \int d^4x\, d^{56}Y\,e\, \Big(& \widehat{R}
 +\frac{1}{48}\,g^{\mu\nu}\,{\cal D}_{\mu}{\cal M}^{MN}\,{\cal D}_{\nu}{\cal M}_{MN}\\
 &{}-\frac{1}{8}\,{\cal M}_{MN}\,{\cal F}^{\mu\nu M}{\cal F}_{\mu\nu}{}^N
 +e^{-1}{\cal L}_{\rm top}-V({\cal M}_{MN},g_{\mu\nu})\Big) \,.
\end{split}
\eea
The theory takes the same structural form as gauged ${\cal N}=8$ supergravity in $D=4$~\cite{deWit:1982ig,deWit:2007mt}, with a
(covariantized) Einstein-Hilbert term for the vierbein $e_{\mu}{}^{a}$, a kinetic term for ${\cal M}$ given by a
non-linear (gauged) sigma-model with target space E$_{7(7)}/$SU$(8)$, a Yang-Mills-type 
kinetic term for the gauge vectors and a `potential' $V({\cal M},g)$ that is a manifestly E$_{7(7)}$ covariant 
expression based only on internal derivatives $\partial_M$. 
In addition, there is a topological Chern-Simons-like term, which is required for consistency with the 
duality relations (\ref{duality56}). 
We stress that here all fields depend on the $4+56$ coordinates, with the internal 
derivatives  entering the non-abelian gauge structure of covariant derivatives and field strengths,
and that the theory encodes in particular $D=11$ supergravity for a particular solution of the 
constraints (\ref{sectioncondition}). 
The detailed construction of all terms in the action will be given below.

 The EFT is uniquely determined by its bosonic gauge symmetries, which are the generalized 
 diffeomorphisms in the external and internal coordinates. 
 In the rest of the introduction we will briefly explain the novel features of its gauge structure. 
As in DFT, the generalized internal diffeomorphisms take the form of generalized Lie derivatives 
$\mathbb{L}_{\Lambda}$ with respect to a vector parameter $\Lambda^M$, e.g., 
$\delta_{\Lambda}{\cal M}_{MN}=\mathbb{L}_{\Lambda}{\cal M}_{MN}$. 
These generalized Lie derivatives, which preserve the E$_{7(7)}$  group 
properties of ${\cal M}_{MN}$, form an algebra according to
 \be
 \label{algIntro}
  \big[\mathbb{L}_{\Lambda_1},\mathbb{L}_{\Lambda_2}\big] \ = \ \mathbb{L}_{[\Lambda_1,\Lambda_2]_{\rm E}}\;, 
 \ee 
modulo the constraints (\ref{sectioncondition}), 
and with the E$_{7(7)}$ E-bracket ${[\Lambda_1,\Lambda_2]_{\rm E}}$ defined by
\bea\label{EbracketIntro}
\big[\Lambda_1,\Lambda_2\big]^M_{\rm E} \ = \ 
2\Lambda_{[1}^K \partial_K \Lambda_{2]}^M
+12 \,(t_\alpha)^{MN}(t^\alpha)_{KL}\,\Lambda_{[1}^K \partial_N \Lambda_{2]}^L
-\frac{1}{4}\Omega^{MN}\Omega_{KL}\partial_N\big(\Lambda_1^K\Lambda_2^L\big)
\;.
\eea
This is the E$_{7(7)}$-covariant extension of the usual Lie bracket in differential geometry. 
However, it does not define a proper Lie algebra in that the Jacobi identity is violated. 
In order to resolve the apparent contradiction with the fact that the 
Lie derivatives define symmetry variations $\delta_{\Lambda}$ of the theory 
(which do satisfy the Jacobi identities), the usual explanation is common to DFT
and the higher-dimensional versions of EFT: the section constraints (\ref{sectioncondition})
imply the existence of gauge parameters that are trivial in the sense that their action on an
arbitrary field vanishes on the `constraint surface' of  (\ref{sectioncondition}).
Specifically, this is the case for gauge parameters given by total (internal) derivatives according to
\bea
\Lambda^M \ \equiv \ (t^\alpha)^{MN}\partial_N \chi_\alpha\;, \qquad {\rm or}\qquad  
\Lambda^M \ \equiv \ \Omega^{MN}\partial_N\chi 
\;, 
\label{trivialIntro}
\eea
with arbitrary $\chi_{\alpha}$ and $\chi$. 
As will become important shortly, however,   
for the E$_{7(7)}$ generalized Lie derivative 
there is actually a more general class of trivial parameters, for which 
there is no direct analogue in DFT or the E$_{6(6)}$ EFT.  These are of the form 
 \be\label{constrainedpara}
  \Lambda^M \ \equiv \ \Omega^{MN}\chi_N\;, \qquad\mbox{with}\;\; \chi_N\;\text{ covariantly constrained}\;, 
 \ee 
where by `covariantly constrained' we denote a field $\chi_M$ that satisfies the same covariant 
constraints (\ref{sectioncondition}) as the internal derivative $\partial_M$, i.e., 
 \be\label{covconstraints}
     (t_\alpha)^{MN}\,\chi_M\, \partial_N  \ = \ (t_\alpha)^{MN}\,\chi_M\, \chi_N \ = \ 0\;, \qquad 
      \Omega^{MN} \,\chi_M\; \partial_N \ = \ 0\;, \quad {\rm etc.}\;, 
  \ee    
in arbitrary combinations and acting on arbitrary functions. 
It is straightforward to see that with $\chi_M=\partial_M\chi$ the class of trivial gauge 
parameters (\ref{constrainedpara}) contains the last term in~(\ref{trivialIntro}) as a special case,
but in general this constitutes a larger class 
which will prove important in the following.
In particular, the Jacobiator associated with (\ref{EbracketIntro}) can be shown to be of the form
 \be\label{JacobIntro}
  J^M(\Lambda_1,\Lambda_2,\Lambda_3) \ \equiv \ 
  3\,\big[\big[\Lambda_{[1},\Lambda_{2}\big]_{\rm E}, \Lambda_{3]}\big]_{\rm E}^M\\
  \ = \ (t^\alpha)^{MN}\partial_N \chi_\alpha(\Lambda)+\Omega^{MN}\chi_N(\Lambda)\;, 
 \ee
where 
 \be
 \begin{split}
  \chi_{\alpha}(\Lambda) \ &= \ -\frac{1}{2}\,(t_{\alpha})_{PQ}\Lambda_1^P[\Lambda_2,\Lambda_3]_{\rm E}^{Q}+{\rm cycl.}\;, \\
  \chi_N(\Lambda) \ &= \ \frac{1}{12}\,\Omega_{PQ}\big(
  \Lambda_1^P\partial_N[\Lambda_2,\Lambda_3]_{\rm E}^Q+[\Lambda_2,\Lambda_3]_{\rm E}^P\,\partial_N\Lambda_1^Q
  +{\rm cycl}.\big)\;,
  \end{split}
  \label{Jacterms}
  \ee 
constitute trivial gauge parameters of the type (\ref{trivialIntro}), (\ref{constrainedpara}).
Thus the Jacobiator has trivial action on all fields and becomes consistent with the 
Jacobi identity for the symmetry variations. Let us stress that the general class
(\ref{constrainedpara}) of trivial gauge parameters is crucial in order to establish consistency of the 
gauge transformations with the Jacobi identity. This seemingly innocent generalization of (\ref{trivialIntro}) has
direct consequences for the required field content and couplings of the theory.

In EFT the gauge transformations, given by generalized Lie derivatives (\ref{algIntro}),
are local both w.r.t.~the internal and external space, i.e., the gauge parameters 
are functions of $x$ and $Y$, $\Lambda^M=\Lambda^M(x,Y)$. 
All external derivatives $\partial_{\mu}$ thus require covariantization by
introduction of an associated gauge connection $A_{\mu}{}^{M}$. We are then faced
with the need to construct a gauge covariant field strength associated to symmetry transformations 
with non-vanishing Jacobiator (\ref{JacobIntro}). This is a standard scenario in the tensor
hierarchy of gauged supergravity~\cite{deWit:2005hv,deWit:2008ta} and solved by 
introducing as compensator fields an appropriate set of 2-form potentials 
with their associated tensor gauge transformations. Applied to our case, the full covariant field strength reads 
  \bea
{\cal F}_{\mu\nu}{}^M &\equiv&
F_{\mu\nu}{}^M - 12 \,  (t^\alpha)^{MN} \partial_N B_{\mu\nu\,\alpha}
-\frac12\,\Omega^{MN} B_{\mu\nu\,N}
\;,
\label{IntromodF7}
\eea
where $F_{\mu\nu}{}^M$ denotes the standard non-abelian Yang-Mills field strength associated with (\ref{EbracketIntro}),
and the 2-forms $B_{\mu\nu\,\alpha}$, $B_{\mu\nu\,M}$ enter in correspondence with the two terms in the Jacobiator (\ref{JacobIntro}). 
The novelty in this field strength, as compared to the corresponding field strength of DFT \cite{Hohm:2013nja} 
and the E$_{6(6)}$ EFT \cite{Hohm:2013vpa},  
is the last term which carries a 2-form $B_{\mu\nu\,M}$
that itself is a covariantly  constrained field in the sense of (\ref{covconstraints}). 
The form of the Jacobiator (\ref{Jacterms}) shows that gauge covariance of the field strength requires this type
of coupling, whereas a (more conventional but weaker) 
compensating term of the form $\Omega^{MN}\partial_NB_{\mu\nu}$ with an unconstrained 
singlet 2-form $B_{\mu\nu}$ would not be sufficient to absorb all non-covariant terms in the variation.

While the notion of such a constrained compensator field may appear somewhat outlandish, 
the above discussion shows that its presence is a direct consequence of the properties of the 
E-bracket Jacobiator for E$_{7(7)}$. 
In turn, this compensator field will play a crucial role in identifying the dynamics of (\ref{duality56}), (\ref{finalaction}),
with the one of the full $D=11$ supergravity. It ensures the correct and duality covariant description of those degrees
of freedom that are on-shell dual to the eleven-dimensional graviton. 
More specifically, after explicit solution of the section constraint (\ref{sectioncondition}) and 
upon matching the field content (\ref{fieldcontent})
with that of $D=11$ supergravity, 7 components among the
56 gauge fields $A_{\mu}{}^{M}$ find their origin in the Kaluza-Klein vectors descending from the $D=11$
metric. The twisted self-duality equations (\ref{duality56}) thus seem to provide a first-order description of 
(at least a part of) the higher-dimensional gravitational dynamics by relating the 7 Kaluza-Klein vectors to 
7 vector fields descending from what should be considered the $D=11$ dual 
graviton~\cite{Curtright:1980yk,Hull:2000zn,West:2001as,Hull:2001iu}. 
Such a duality is commonly recognized to be restricted to the linearized level on the grounds of the no-go results
of~\cite{Bekaert:2002uh,Bekaert:2004dz}. 
The non-linear equations (\ref{duality56}) circumvent this problem precisely by virtue of the
covariantly constrained compensator fields $B_{\mu\nu\,M}$, which can be viewed as a 
covariantization of the formulation of~\cite{Boulanger:2008nd}.
As a result, the E$_{7(7)}$-covariant model (\ref{duality56}), (\ref{finalaction}), upon appropriate solution of
the section constraint (\ref{sectioncondition}), precisely reproduces the complete
set of untruncated $D=11$ field equations while featuring components of the dual graviton.
The very same pattern has been observed in the $3D$ duality-covariant formulation of $D=4$ Einstein gravity in~\cite{Hohm:2013jma}
where the constrained compensator gauge fields appear among the gauge vectors. 
In contrast, in the ${\rm E}_{6(6)}$-covariant construction of~\cite{Hohm:2013vpa}, the degrees of freedom from the higher-dimensional
dual graviton do not figure among the fields in the EFT action and the constrained compensator fields only enter the $p$-form hierarchy
at the level of the three-forms.

We finally note that while 
the above action (\ref{finalaction}) is manifestly invariant under the internal generalized diffeomorphisms 
with gauge parameter   $\Lambda^M$ (in the sense that each term is separately invariant), it also features a 
non-manifest gauge invariance under diffeomorphisms in the external coordinates $x^{\mu}$ (with the parameter $\xi^\mu$
depending on coordinates $x$ and $Y$). In fact, it is this symmetry, to be discussed below in more detail, 
that determines all relative coefficients in (\ref{finalaction}).  

This paper is organized as follows. In sec.~\ref{sec2} we introduce the details of the E$_{7(7)}$ 
generalized Lie derivatives and its E-bracket algebra, together with the associated 
covariant derivatives, field strengths and the tensor hierarchy. With these ingredients at 
hand, we define in sec.~\ref{sec3} the full E$_{7(7)}$  EFT, including a discussion 
of the non-manifest invariance under $(3+1)$-dimensional diffeomorphism of the $x^{\mu}$. 
In sec.~4 we discuss the embedding of 11-dimensional supergravity and IIB supergravity upon choosing
particular solutions of the section constraint. We conclude in sec.~\ref{sec:5}, while we collect 
some important   E$_{7(7)}$ relations in the appendix.

\section{E$_{7(7)}$ Generalized Diffeomorphisms and the Tensor Hierarchy}\label{sec2}

In this section, we introduce the E$_{7(7)}$ generalized Lie derivatives that 
generate the internal (generalized) diffeomorphisms and the E-bracket and work out the
associated tensor hierarchy. Vector fields $A_{\mu}{}^{M}$ in the fundamental 56-dimensional 
representation of E$_{7(7)}$ act as gauge fields in order to covariantize the theory under
$x$-dependent internal (generalized) diffeomorphisms. The non-trivial Jacobiator of the E-bracket
further requires the introduction of the two-form $B_{\mu\nu\, \alpha}$ in the adjoint of E$_{7(7)}$  
in accordance with the general tensor hierarchy of non-abelian $p$-forms~\cite{deWit:2005hv,deWit:2008ta}.
Up to this point, the construction is completely parallel to the construction of the ${\rm E}_{6(6)}$-covariant tensor hierarchy,
presented in detail in~\cite{Hohm:2013vpa}. We will thus keep the presentation brief and compact. 
The new ingredient w.r.t.\ to the ${\rm E}_{6(6)}$-covariant construction  
is the appearance of a covariantly constrained compensating gauge field $B_{\mu\nu\, M}$ among the two-forms,
whose presence is required by closure of the tensor hierarchy.
This field takes values in the fundamental representation of E$_{7(7)}$, however, restricted by
covariant constraints, see (\ref{constraintsB}) below.

\subsubsection*{Generalized Lie derivative and E-bracket}
Let us start by collecting the relevant ingredients of the exceptional Lie group E$_{7(7)}$. 
Its Lie algebra is of dimension 133, with generators that we denote by  
$t_{\alpha}$ with the adjoint index $\alpha=1,\ldots, 133$. The fundamental representation of 
E$_{7(7)}$ is of dimension 56 and denoted by indices $M,N=1,\ldots,56$. The symplectic embedding
E$_{7(7)}\subset {\rm Sp}(56)$ implies the existence of an invariant antisymmetric tensor $\Omega^{MN}$
which we will use to raise and lower fundamental indices, adopting north-west south-east conventions:
$V^M=\Omega^{MN} V_N$, $V_M=V^N\Omega_{NM}$, with $\Omega^{MK}\Omega_{NK} = \delta_{N}{}^{M}$. 
In contrast, adjoint indices are raised and lowered by the (rescaled) symmetric Cartan-Killing form
$\kappa_{\alpha\beta}\equiv (t_{\alpha})_M{}^N (t_{\beta})_N{}^M$\,.
Due to the invariance of $\Omega^{MN}$, the gauge group generator in the fundamental representation with 
one index lowered, $(t_{\alpha})_{MN}$, is \textit{symmetric} in its two fundamental indices.   
Below we will need the projector onto the adjoint representation 
\bea\label{adjproj}
\mathbb{P}^K{}_M{}^L{}_N&\equiv&
(t_\alpha)_M{}^K (t^\alpha)_N{}^L \nonumber\\
&=&
\frac1{24}\,\delta_M^K\delta_N^L
+\frac1{12}\,\delta_M^L\delta_N^K
+(t_\alpha)_{MN} (t^\alpha)^{KL}
-\frac1{24} \,\Omega_{MN} \Omega^{KL}
\;, 
\eea
which satisfies 
\bea
\mathbb{P}^M{}_N{}^N{}_M &=& 133\;.
\eea

Next, we introduce the generalized Lie derivative w.r.t.~the vector parameter $\Lambda^M$. 
Its action on a vector $V^M$ of  weight $\lambda$ is defined as \cite{Coimbra:2011ky,Berman:2012vc}
\bea\label{genLie}
\delta V^M \ = \ \mathbb{L}_{\Lambda} V^M 
\ \equiv \ \Lambda^K \partial_K V^M - 12\, \mathbb{P}^M{}_N{}^K{}_L\,\partial_K \Lambda^L\,V^N
+\lambda\,\partial_P \Lambda^P\,V^M
\;,
\eea
with appropriate generalization for its action on an E$_{7(7)}$ tensor 
with an arbitrary number of fundamental indices.
Because of the projector in (\ref{genLie}), the generalized Lie derivative is compatible with the ${\rm E}_{7(7)}$ algebra
structure: e.g.\ the $\Omega$-tensor is an invariant tensor of weight 0
\bea
\mathbb{L}_{\Lambda}\,\Omega^{MN} &=& 0
\;,
\eea
implying that the definition (\ref{genLie}) also induces the proper covariant transformation behavior for the covariant
vector $V_M\equiv\Omega_{NM} V^N$\,.
Explicitly, writing out the projector (\ref{adjproj}), the Lie derivative (\ref{genLie}) reads 
\bea\label{explicitLie}
   \delta _{\Lambda}V^M &=&   \Lambda^K \partial_K V^M-\partial_N\Lambda^M V^N
  +\Big(\lambda-\frac{1}{2}\Big)\,\partial_P\Lambda^P\, V^M
 \nonumber\\
 &&{}
 -12\,
  (t_\alpha)^{MN} (t^\alpha)_{KL}\,\partial_N \Lambda^K V^L 
  -\frac12\,\Omega^{MN}\Omega_{KL}\,\partial_N \Lambda^K V^L
\;. 
\eea

We now discuss some properties of the generalized Lie derivative. As mentioned in the introduction, 
there are trivial gauge parameters that do not generate a gauge transformation. They are  of the form 
\bea
\Lambda^M \ \equiv \ (t^\alpha)^{MN}\partial_N \chi_\alpha\;, \qquad 
\Lambda^M \ = \  \Omega^{MN}\chi_{N} 
\;, 
\label{trivial}
\eea
with a covariantly constrained co-vector $\chi_M$ in the sense of satisfying (\ref{covconstraints}). 
In order to state the constraints in more compact form, let us introduce the projector $\mathbb{P}_{{\bf{1+133}}}$
onto the ${\bf 1}\oplus {\bf 133}$ sub-representation 
in the tensor product ${\bf 56}\otimes {\bf 56}$.
In terms of this projector  the constraints (\ref{covconstraints}) take the compact form  
 \be\label{moretrivial}
  (\mathbb{P}_{{\bf{1+133}}})^{MN}\,\chi_M\,\partial_N \ = \ 0 \ = \
   (\mathbb{P}_{{\bf{1+133}}})^{MN}\, \chi_M\,\chi_N \;.
 \ee 
The triviality of  $\Lambda^M = \Omega^{MN}\chi_{N}$ follows by a straightforward 
explicit calculation, using the identity (\ref{A1}) and making repeated use of the constraints. 
The triviality of the first parameter in (\ref{trivial}) follows similarly by a straightforward but somewhat 
more involved computation, using the identities in the appendix.

Let us now discuss the algebra of gauge transformations (\ref{genLie}). 
A direct computation making use of the algebraic identities collected in appendix~\ref{app:relations} 
shows that modulo the section constraints (\ref{sectioncondition}), these gauge transformations 
close~\cite{Coimbra:2011ky,Berman:2012vc}, 
 \bea\label{gaugeclosure}
{}\big[ \delta_{\Lambda_1},\delta_{\Lambda_2}\big] \ = \ \delta_{[\Lambda_2,\Lambda_1]_{\rm E}}\;,
\eea
according to   the `E-bracket' 
\bea\label{Ebracket}
\big[\Lambda_2,\Lambda_1\big]^M_{\rm E} \ = \ 
2\Lambda_{[2}^K \partial_K \Lambda_{1]}^M
+12 \,(t_\alpha)^{MN}(t^\alpha)_{KL}\,\Lambda_{[2}^K \partial_N \Lambda_{1]}^L
-\frac{1}{4}\Omega^{MK}\Omega_{NL}\partial_K\big(\Lambda_2^N\Lambda_1^L\big)
\;.
\eea
Note that the last term in here is actually of the trivial form (\ref{trivial}) and so does not 
generate a gauge transformation. This term is therefore ambiguous, and the reason we added it 
here (with this particular coefficient) is that the associated Jacobiator, 
i.e.~the failure of the E-bracket to satisfy the Jacobi identity,  
takes a simple form. The appearance of this term is novel compared to the E$_{6(6)}$ case
and therefore we go in some detail through the proof of the triviality of the Jacobiator. 
We first need some notation and define the Dorfman-type product between vectors of weight $\frac{1}{2}$ as
 \be
 \begin{split}
 (V\circ W)^M \ \equiv \ (\mathbb{L}_{V}W)^M 
 \ = \ &
 V^K \partial_K W^M-W^K \partial_K V^M\\
& -12\,
  (t_\alpha)^{MN} (t^\alpha)_{KL}\,\partial_N V^K W^L 
  -\frac12\,\Omega^{MN}\Omega_{KL}\,\partial_N V^K W^L\;.
 \end{split} 
 \ee 
Comparing this with the E-bracket we conclude
 \bea\label{EDdifference}
   (V\circ W)^M &=& \big[V,W\big]_{\rm E}^M
   -6(t^\alpha)^{MN} \partial_N\big((t_{\alpha})_{KL}W^KV^L\big)
   +\frac{1}{4}\Omega^{MK}\Omega_{NL}\big(V^N\partial_KW^L
   +W^N\partial_KV^L\big) \nonumber\\
    &\equiv &  \big[V,W\big]_{\rm E}^M+\big\{ V,W\big\}^M
   \;, 
 \eea
introducing for later convenience the short-hand notation in the second line defined by the symmetric 
pairing in the first equation.    
In contrast to the situation in DFT and the E$_{6(6)}$ E-bracket, the final term in the first line 
cannot be written as a total derivative.   
Rather, it is of a trivial form in the stronger sense of (\ref{moretrivial}). 
Therefore, both terms generate a trivial action, 
and we have 
 \be\label{trivialLie} 
  \mathbb{L}_{[V,W]_{\rm E}} \ = \ \mathbb{L}_{(V\circ W)}\;.
 \ee
Another important property is that the antisymmetrized Dorfman product 
coincides with the E-bracket as defined in (\ref{Ebracket}), 
 \be\label{symmprop}
  \frac{1}{2}(V\circ W-W\circ V) \ = \ \big[V,W\big]_{\rm E}\;. 
 \ee 
It is this property that determines the a priori ambiguous coefficient of the  $\Omega\Omega$ term  
in the E-bracket. 
Finally, the Dorfman product satisfies the Jacobi-like (or Leibniz-type) identity 
 \be\label{Leibniz}
  U\circ(V\circ W) \ = \ (U\circ V)\circ W+V\circ(U\circ W)\;.
 \ee  
This follows from the algebra and the property (\ref{trivialLie}) in complete analogy to the discussion 
in \cite{Hohm:2013vpa}. 
It is now straightforward to compute the Jacobiator 
 \be\label{JacDEF}
  J(V_1,V_2,V_3) \ \equiv \ 3\big[\big[V_{[1},V_{2}\big]_{\rm E}, V_{3]}\big]_{\rm E}
  \ = \ -3\big[V_{[1},\big[V_2,V_{3]}\big]_{\rm E}\big]_{\rm E}\;. 
 \ee 
In the following computation we will assume total antisymmetrization in the three arguments 
$1,2,3$, but not display it explicitly. Keeping this in mind we compute for the term on the right-hand side
with (\ref{symmprop}) and (\ref{Leibniz}):
 \be\label{Jac1}
 \begin{split}
  \big[V_{1},\big[V_2,V_{3}\big]_{\rm E}\big]_{\rm E} \ &= \   \big[V_{1},V_2\circ V_3\big]_{\rm E}
  \ = \ \frac{1}{2}\big(V_1\circ(V_2\circ V_3) -(V_2\circ V_3)\circ V_1\big)\\
  \ &= \  \frac{1}{2}\big((V_1\circ V_2)\circ V_3+V_2\circ(V_1\circ V_3) -(V_2\circ V_3)\circ V_1\big)\\
  \ &= \ -\frac{1}{2}V_1\circ(V_2\circ V_3)\;, 
 \end{split} 
 \ee 
where we recalled the total antisymmetry in the last step. Thus, the E-bracket Jacobiator 
is proportional to the `Dorfman-Jacobiator'. On the other hand, 
from (\ref{EDdifference}) we also have 
 \be
   \big[V_{1},\big[V_2,V_{3}\big]_{\rm E}\big]_{\rm E} \ = \   \big[V_{1},V_2\circ V_3\big]_{\rm E}
   \ = \ V_1\circ(V_2\circ V_3)-\big\{ V_1, \big[V_2, V_3 \big]_{\rm E}\big\}\;. 
 \ee  
Using that this equals (\ref{Jac1}) we can determine the Dorfman-Jacobiator
and, via (\ref{Jac1}) again, the E-bracket Jacobiator (\ref{JacDEF}), 
 \be
  J(V_1,V_2,V_3) 
  \ = \ \frac{1}{3}\Big(\big\{ V_1, [V_2, V_3 ]_{\rm E}\big\}
  +\big\{ V_2, [V_3, V_1 ]_{\rm E}\big\}+\big\{ V_3, [V_1, V_2 ]_{\rm E}\big\}\Big)\;, 
 \ee
writing out the total antisymmetrization. This shows that the Jacobiator is of a trivial 
form that does not generate a gauge transformation. 
More explicitly, using the notation introduced in (\ref{EDdifference}), the Jacobiator is 
 given by
  \bea\label{JacobiatorFinal}
  J^M(V_1,V_2,V_3) &=&-\frac{1}{2}\,(t_\alpha)^{MK}\partial_K
  \Big((t^\alpha)_{PL}\big(\,V_1^P[V_2,V_3]_{\rm E}^L+{\rm cycl.}\big)\Big)
  \nonumber\\
  &&{}+\frac{1}{12}\Omega^{MK}\Omega_{NL}\big(V_1^N\partial_K[V_2,V_3]_{\rm E}^L
  +[V_1,V_2]_{\rm E}^N\, \partial_K V_3^L+{\rm cycl.}\big)
  \;.
 \eea

So far, we have discussed the action of the generalized Lie derivative on vectors in
the fundamental representation of ${\rm E}_{7(7)}$.
From (\ref{genLie}), we likewise obtain the action of the Lie derivative on a tensor 
in the adjoint representation (of weight $\lambda'$)
\bea
\delta W_\alpha &=& 
\Lambda^K \partial_K W_\alpha 
+ 12 \,f_{\alpha\beta}{}^\gamma\,(t^\beta)_L{}^K\,\partial_K\Lambda^L\,W_\gamma
+\lambda'\,\partial_K\Lambda^K\,W_\alpha
\;,
\label{Liealpha}
\eea
with the ${\rm E}_{7(7)}$ structure constants $f_{\alpha\beta}{}^\gamma$\,.
By construction, the ${\rm E}_{7(7)}$ generators $(t_\alpha)^{MN}$ then are
invariant tensors of weight 0 w.r.t.\ the generalized Lie derivative.
In the following 
 we will be led to consider such adjoint tensors under internal derivatives, 
more specifically combinations of the type
\bea
T^M &\equiv& 
 (t^\alpha)^{MN}\,\partial_N W_\alpha
 \;.
 \label{VV}
\eea
Some straightforward computation (and use of some of the algebraic relations collected
in appendix~\ref{app:relations}) shows that under the generalized Lie derivative, the combination (\ref{VV})
transforms as
\bea
\delta_\Lambda\, T^M &=&
\Lambda^K \partial_K  T^M - 12\,{\mathbb{P}}^M{}_N{}^K{}_L\,\partial_K\Lambda^L\, T^N
+\Big(\lambda'-\frac12\Big)\,\partial_K\Lambda^K\, T^M
\nonumber\\
&&{}
+(\lambda'-1)\,(t^\alpha)^{MN} W_\alpha \,\partial_N \partial_K\Lambda^K
+  \Omega^{MN}  (t^\alpha)_L{}^K \,W_\alpha
\,\partial_N \partial_K\Lambda^L
\;.
\label{deltaT}
\eea
The first line amounts to the covariant transformation of a vector of weight $\lambda=\lambda'-\frac12$, 
while the second line represents non-covariant terms. 
The full result (\ref{deltaT}) then shows that for $\lambda'=1$, $T^M$ transforms like a contravariant 
vector of weight $\lambda=\frac12$ up to a term proportional to $\Omega^{MN}\partial_N$.
To correct for the latter, we may introduce a compensating field $W_M$ subject to the same 
constraints as those discussed in (\ref{moretrivial}), i.e.
 \be\label{constraintsW}
  (\mathbb{P}_{{\bf{1+133}}})^{MN}\,W_M\,\partial_N \ = \ 0 \ = \ 
   (\mathbb{P}_{{\bf{1+133}}})^{MN}\, W_M\,W_N \;,
 \ee 
and consider the combination
\bea
\widehat{T}^M &\equiv& (t^\alpha)^{MN}\,\partial_N W_\alpha 
+ \frac1{24}\,\Omega^{MN}\,W_N
\;.
\label{tensorT}
\eea
This combination then transforms as a covariant vector of weight $\lambda=\frac12$, 
\bea
\delta_\Lambda\,\widehat{T}^M &=& 
\Lambda^K \partial_K  \widehat{T}^M 
- 12\,{\mathbb{P}}^M{}_N{}^K{}_L\,\partial_K\Lambda^L\, \widehat{T}^N
+\frac12\,\partial_K\Lambda^K\, \widehat{T}^M
\;,
\label{deltaVhat}
\eea
provided the compensating field $W_M$ transforms as
\be
\delta_\Lambda W_M =
\Lambda^K \partial_K  W_M 
+ 12\,{\mathbb{P}}^N{}_M{}^K{}_L\,\partial_K\Lambda^L\, W_N
+\frac12\,\partial_K\Lambda^K\, W_M
-24\,(t^\alpha)_L{}^K \,W_\alpha \,\partial_M \partial_K\Lambda^L
\;.
\label{deltaLcomp}
\ee
A short calculation confirms that the transformation (\ref{deltaLcomp}) indeed 
preserves the constraints (\ref{constraintsW}) on $W_{M}$\,.
The tensorial nature of (\ref{tensorT}) will prove crucial below for the structure of the tensor hierarchy of non-abelian $p$-forms.  
We note that this crucially hinges on the introduction of the compensating field $W_M$.

\subsubsection*{Covariant derivatives and tensor hierarchy}

We will now introduce gauge connections $A_\mu{}^M$ which manifestly render the model invariant
 under generalized Lie derivatives (\ref{genLie}) with $x$-dependent gauge parameters $\Lambda^M$, 
 covariantizing the derivatives in the usual fashion, 
 \be
  \partial_{\mu}\;\rightarrow\; {\cal D}_{\mu} \ \equiv \ \partial_{\mu}-\mathbb{L}_{A_{\mu}}\;.
 \ee   
Explicitly, from (\ref{explicitLie}) we infer the form of the covariant derivative of a vector of weight $\lambda$, 
\bea
  {\cal D}_\mu V^M  &\equiv& {D}_\mu V^M - \lambda \,\partial_K A_\mu {}^K\,V^M\nonumber\\[1ex]
  &\equiv&
   \partial_\mu V^M-A_\mu {}^K\partial_KV^M+V^K\partial_K A_\mu {}^M+\frac{1-2\lambda}{2}\, 
  \partial_K A_\mu {}^K V^M\nonumber\\
  &&{}+12\,
  (t_\alpha)^{MN} (t^\alpha)_{KL}\,\partial_N A_\mu {}^K V^L 
  +\frac12\,\Omega^{MN}\Omega_{KL}\,\partial_N A_\mu {}^K V^L
  \;.
  \label{covDV}
 \eea
The gauge variation of the vector field $A_\mu{}^M$ is obtained by requiring that the covariant derivative transforms covariantly,  which imposes
\bea
\delta A_\mu{}^M &=& \partial_\mu\Lambda^M - A_\mu{}^K\partial_K \Lambda^M + \Lambda^K \partial_K A_\mu{}^M 
\nonumber\\
&&{}
+12\,(t_\alpha)^{MN} (t^\alpha)_{KL}\,\Lambda^L\, \partial_N A_\mu{}^K
  +\frac12\,\Omega^{MN}\Omega_{KL}\,\Lambda^L \, \partial_N A_\mu{}^K
\nonumber\\[0.3ex]
&=& D_\mu \Lambda^M -\frac12\,(\partial_K A_\mu{}^K)\,\Lambda^M ~\equiv~ {\cal D}_\mu \Lambda^M
\;,
\label{deltaA}
\eea
showing that the gauge parameter $\Lambda^M$ is a  tensor of weight $\lambda=\frac12$.
The associated Yang-Mills field strength, 
\bea
F_{\mu\nu}{}^M &\equiv & 2 \partial_{[\mu} A_{\nu]}{}^M-\big[A_{\mu},A_{\nu}\big]^M_{\rm E} \\ \nonumber
 \ &=& \ 2 \partial_{[\mu} A_{\nu]}{}^M 
-2\,A_{[\mu}{}^K \partial_K A_{\nu]}{}^M 
-\frac1{2}\left(24\, (t_\alpha)^{MK} (t^\alpha)_{NL}
-\Omega^{MK}\Omega_{NL}\right)\,A_{[\mu}{}^N\,\partial_K A_{\nu]}{}^L
\,,
\nonumber
\label{YM}
\eea
has a general variation given by
\bea
\delta F_{\mu\nu}{}^M 
&=&
2 D_{[\mu} \delta A_{\nu]}{}^M - \partial_K A_{[\mu}{}^K\,\delta A_{\nu]}{}^M
-12\, (t_\alpha)^{MK}(t^\alpha)_{NL}\, \partial_K\left(  A_{[\mu}{}^N\,\delta A_{\nu]}{}^L\right)
\nonumber\\
&&{}
-\frac12\,\Omega^{MK}\Omega_{LN}\,
\left( A_{[\mu}{}^N\, \partial_K \delta A_{\nu]}{}^L
-\partial_K A_{[\mu}{}^N \delta A_{\nu]}{}^L \right)
\;,
\label{deltaFYM}
\eea
and is not covariant w.r.t.~vector gauge transformations (\ref{deltaA}).
This is a consequence of the non-vanishing Jacobiator~(\ref{JacobiatorFinal}).
In order to define a covariant field strength, it is natural in the spirit of the tensor hierarchy~\cite{deWit:2005hv,deWit:2008ta}
to extend the field strength (\ref{YM}) by further St\"uckelberg type couplings according to
\bea
{\cal F}^\circ_{\mu\nu}{}^M &\equiv&
F_{\mu\nu}{}^M - 12 \,  (t^\alpha)^{MN} \,\partial_N B_{\mu\nu\,\alpha}
\;,
\label{modF0}
\eea
to two-form tensors 
$B_{\mu\nu\,\alpha}$ in the adjoint representation of E$_{7(7)}$, whose transformations may absorb some
of the non-covariant terms in (\ref{deltaFYM}). However, unlike the E$_{6(6)}$ covariant construction
of \cite{Hohm:2013vpa}, this modification is not sufficient in order to obtain fully gauge covariant field strengths.
In particular, the last line of (\ref{deltaFYM}) continues to spoil the proper transformation behavior of the field strength
and cannot be absorbed into a transformation of $B_{\mu\nu\,\alpha}$. 
This indicates  that in the E$_{7(7)}$ covariant construction new fields are required at the level
of the two-form tensors, as discussed in the introduction.  
We recall that with five external dimensions, these additional fields only enter at the level of the three-forms
and remain invisible in the action~\cite{Hohm:2013vpa}, whereas in the three-dimensional case they 
are already present among the vector fields~\cite{Hohm:2013jma}.
The fully covariantized field strength is given by the expression
\bea
{\cal F}_{\mu\nu}{}^M &\equiv&
F_{\mu\nu}{}^M - 12 \,  (t^\alpha)^{MN} \,\partial_N B_{\mu\nu\,\alpha}
-\frac12\,\Omega^{MK}\,B_{\mu\nu\,K}
\;,
\label{modF7}
\eea
where the two-form $B_{\mu\nu\,K}$ is a covariantly constrained compensating gauge field, 
i.e.\ a field subject to the same section constraints as the internal derivatives, 
\bea
(\mathbb{P}_{{\bf 1}+{\bf 133}})^{MN}\,B_M \partial_N  &=& 0\;,\qquad
(\mathbb{P}_{{\bf 1}+{\bf 133}})^{MN}\, B_M B_N ~=~ 0 \;.
\label{constraintsB}
\eea 
The general variation of ${\cal F}_{\mu\nu}{}^M$ is given by
\bea
\delta {\cal F}_{\mu\nu}{}^M &=&
2\,{\cal D}_{[\mu} \delta A_{\nu]}{}^M 
- 12 \,  (t^\alpha)^{MN} \,\partial_N \Delta B_{\mu\nu\,\alpha}
-\frac12\,\Omega^{MK}\,\Delta B_{\mu\nu\,K}
\;,
\eea
with the E$_{7(7)}$ tensor $\delta A_\mu{}^M$ of weight $\lambda=\frac12$, and
\bea
\Delta B_{\mu\nu\,\alpha} &\equiv& \delta B_{\mu\nu\,\alpha} +  
(t_\alpha)_{KL}\,A_{[\mu}{}^K\, \delta A_{\nu]}{}^L\;,
\nonumber\\
\Delta B_{\mu\nu\,K} &\equiv& \delta B_{\mu\nu\,K}
+\Omega_{LN}\left(A_{[\mu}{}^N \partial_K \delta A_{\nu]}{}^L
-\partial_K A_{[\mu}{}^N \, \delta A_{\nu]}{}^L\right)
\;.
\label{DeltaB}
\eea
In particular, we may define vector gauge variations 
\bea
   \delta_\Lambda A_\mu{}^M &=&  {\cal  D}_\mu \Lambda^M  \;, 
   \nonumber\\
      \Delta_\Lambda B_{\mu\nu \alpha} &=&  
   (t_\alpha)_{KL}\, \Lambda^K{\cal F}_{\mu\nu}{}^{L}\;, 
     \nonumber\\
     \Delta_\Lambda B_{\mu\nu M} &=& - \Omega_{KL}
     \left({\cal F}_{\mu\nu}{}^K\partial_M \Lambda^L-
     \Lambda^L\partial_M {\cal F}_{\mu\nu}{}^K \right)
     \;,
     \label{gaugeL}
 \eea
under which the field strength ${\cal F}_{\mu\nu}{}^M$ transforms covariantly 
\bea
\delta_\Lambda {\cal F}_{\mu\nu}{}^M &=&
\Lambda^K \partial_K {\cal F}_{\mu\nu}{}^M 
- 12 \, \mathbb{P}^M{}_N{}^K{}_L\,\partial_K \Lambda^L\,{\cal F}_{\mu\nu}{}^N
+\frac12\,\partial_K \Lambda^K\,{\cal F}_{\mu\nu}{}^M
\;,
\label{deltaF}
\eea
i.e., as an E$_{7(7)}$  vector of weight $\lambda=\frac12$\,.
As part of this calculation, we have used that
\bea
\mathbb{L}_{F_{\mu\nu}}\,\Lambda^M &=&\mathbb{L}_{{\cal F}_{\mu\nu}}\,\Lambda^M\;,
\eea 
which states that $F_{\mu\nu}$ and ${\cal F}_{\mu\nu}$ differ by terms
that are trivial and so do not generate a generalized Lie derivative, c.f.~(\ref{trivial}).
Let us also note that the form of the gauge transformations (\ref{DeltaB}), (\ref{gaugeL})
manifestly preserves the constraints (\ref{constraintsB}) on the compensating 
gauge field as a consequence of (\ref{sectioncondition}).

The two-form tensors $B_{\mu\nu\,\alpha}$ and $B_{\mu\nu\,M}$ carry their own
gauge symmetries which act as
\bea
   \delta A_\mu{}^M &=& 12\,(t^\alpha)^{MN}\,\partial_N\Xi_{\mu\,\alpha}
   +\frac12\,\Omega^{MN}\,\Xi_{\mu\,N} \;,
   \nonumber\\
      \Delta B_{\mu\nu \alpha} &=& 2\,{\cal D}_{[\mu}\Xi_{\nu]\alpha} \;, 
      \nonumber\\
      \Delta B_{\mu\nu M} &=& 2\,{\cal D}_{[\mu}\Xi_{\nu]M} 
           +48\,(t^\alpha)_L{}^K  \left(\partial_K \partial_M A_{[\mu}{}^L\right) \Xi_{\nu]\alpha}
      \;,
      \label{tensor_gauge7}
 \eea
and leave the field strength (\ref{modF7}) invariant.
The tensor gauge parameters $\Xi_{\mu\,\alpha}$ and $\Xi_{\mu\,M}$
are of weight $\lambda'=1$ and $\lambda=\frac12$, respectively,
with their covariant derivatives defined according to 
(\ref{explicitLie}) and (\ref{Liealpha}), respectively.
Note that the seemingly non-covariant term in  $\Delta B_{\mu\nu M}$  has its origin 
in the final term in (\ref{deltaLcomp}), which reflects that the constrained 
field $B_{M}$ does not have separately tensor character, but only in combinations of the type~(\ref{tensorT}). 
In particular, the computation of invariance of the field strength ${\cal F}_{\mu\nu}{}^M$ under (\ref{tensor_gauge7})
crucially depends on the observation that a tensor combination according to (\ref{tensorT}) is again of tensorial nature.

We close this presentation of the tensor fields by stating the Bianchi identities
 \be
  3\, {\cal D}_{[\mu}{\cal F}_{\nu\rho]}{}^M \ = \ 
 - 12\, (t^\alpha)^{MN}\partial_N{\cal H}_{\mu\nu\rho\,\alpha} - \frac12\,\Omega^{MN}\,
  {\cal H}_{\mu\nu\rho\,N}
  \;,
  \label{Bianchi7}
 \ee 
with the three-form field strengths ${\cal H}_{\mu\nu\rho\,\alpha}$ and ${\cal H}_{\mu\nu\rho\,N}$ defined by this equation up to terms
that vanish under the projection  with $(t^\alpha)^{MN}\partial_N$.
This identity again is a nice illustration of tensorial structures of the type (\ref{tensorT}),
with the field strength ${\cal H}_{\mu\nu\rho\,M}$ 
transforming according to (\ref{deltaLcomp}) under generalized Lie derivatives.

\section{Covariant E$_{7(7)}$ Theory}\label{sec3}

With the tensor hierarchy associated to generalized diffeomorphisms set up,
we are now in the position to define the various terms in the action (\ref{finalaction})
and the duality equation (\ref{duality56}). We then verify that the complete set
of equations of motion is invariant under generalized internal and external diffeomorphisms,
which in turn fixes all the couplings.

\paragraph{Kinetic terms}
The metric, the scalar fields and the vector gauge fields come with second order
kinetic terms in the action (\ref{finalaction}). As in~\cite{Hohm:2013vpa,Hohm:2013nja}, the Einstein-Hilbert term is built from the 
improved Riemann tensor 
  \be
  \widehat{R}_{\mu\nu}{}^{ab} \ \equiv \  R_{\mu\nu}{}^{ab}[\omega]+{\cal F}_{\mu\nu}{}^{M}
  e^{a}{}^{\rho}\partial_M e_{\rho}{}^{b}\,,
  \label{improvedRE7}
 \ee
where $R_{\mu\nu}{}^{ab}[\omega]$ denotes the curvature of the spin connection which in turn is 
given by the standard expression in terms of the vierbein with all
derivatives covariantized according to
 \bea
{\cal D}_\mu e_\nu{}^a &\equiv& \partial_\mu e_\nu{}^a - A_\mu{}^M\partial_M e_\nu{}^a
-\frac12\, \partial_MA_\mu{}^M e_\nu{}^a
\,.
\label{covderE}
\eea
I.e., the vierbein is an ${\rm E}_{7(7)}$ scalar of weight $\lambda=\frac12$\,.
The covariantized Einstein-Hilbert term 
 \be
  {\cal L}_{\rm EH} \ = \ e\,\widehat{R} \ = \ 
  e\,e_{a}{}^{\mu}e_{b}{}^{\nu} \widehat{R}_{\mu\nu}{}^{ab}\;, 
 \ee
then is invariant under Lorentz transformations and
correctly transforms as a density under internal generalized diffeomorphisms
with the weight $2$ of the vierbein determinant and the weights $-\frac{1}{2}$ of the inverse vierbeins adding up to 1.
The 70 scalar fields of the theory parametrize the coset space ${\rm E}_{7(7)}/{\rm SU}(8)$,
which is conveniently described by the symmetric $56\times56$ matrix ${\cal M}_{MN}$, with the kinetic term
given by
\bea
{\cal L}_{\rm sc} &=& \frac1{48}\,e\,g^{\mu\nu}\,{\cal D}_{\mu}{\cal M}_{MN}\,{\cal D}_{\nu}{\cal M}^{MN}
\;,
\label{Lscal}
\eea
with the inverse matrix ${\cal M}^{MN}$ related by
\bea
{\cal M}^{MN} &=& \Omega^{MK}\Omega^{NL} \,{\cal M}_{KL}
\;,
\eea
as a consequence of the symplectic embedding of ${\rm E}_{7(7)}$.
All derivatives in (\ref{Lscal}) are covariantized as (\ref{covDV}) with ${\cal M}_{MN}$ transforming as
an ${\rm E}_{7(7)}$ tensor of weight $\lambda=0$\,. This 
is compatible with the group property ${\rm det}\,{\cal M}_{MN}=1$\,.
As for the Einstein-Hilbert term, the total weight of (\ref{Lscal}) is $1$  
as required for $\Lambda^M$ gauge invariance. 
Finally, also the Yang-Mills kinetic term 
\bea
{\cal L}_{\rm YM}&=&-\frac{1}{8}\,e\, {\cal M}_{MN}\,{\cal F}^{\mu\nu M}{\cal F}_{\mu\nu}{}^{N}
\;,
\label{LYM}
\eea
carries the correct weight of 1, since the field strengths transform as tensors of weight $\lambda=\frac12$,
c.f., (\ref{deltaF}). As discussed above, this term gives rise to second order field equations for all
56 vector fields $A_\mu{}^M$ whereas the Lagrangian (\ref{finalaction}) is amended by the covariant first-order duality equations
\bea
{\cal E}_{\mu\nu}{}^M &\equiv&  
{\cal F}_{\mu\nu}{}^M +\frac12\,e\,\varepsilon_{\mu\nu\rho\sigma}\,\Omega^{MN}\,{\cal M}_{NK}\,{\cal F}^{\rho\sigma\,K}
~=~0
\;,
\label{duality}
\eea
which ensures that only 28 of them correspond to independent propagating degrees of freedom.
Both terms in this duality equation are ${\rm E}_{7(7)}$ tensors of weight $\lambda=\frac12$.
 
\paragraph{Topological term}
 The topological term is required in order to ensure that the variation of the two-form tensors in (\ref{LYM})
 does not give rise to inconsistent field equations. This term is most conveniently constructed as the boundary
 term of a manifestly gauge invariant exact form in five dimension as
 \bea
 \label{Ltop}
 S_{\rm top} &=& -\frac1{24}\,\int_{\Sigma_5} d^5x \int d^{56}Y\,
\varepsilon^{\mu\nu\rho\sigma\tau}\,{\cal F}_{\mu\nu}{}^M\,
{\cal D}_{\rho} {\cal F}_{\sigma\tau}{}_M
\nonumber\\
&\equiv& 
\int_{\partial\Sigma_5} d^4x \int d^{56}Y\,
{\cal L}_{\rm top}
\;.
 \eea
The explicit form of the four-dimensional Lagrangian density is not particularly illuminating, 
since it is not manifestly gauge invariant.
What we will need in the following is its variation
 \bea
\delta {\cal L}_{\rm top} &=&
-\frac14\,\varepsilon^{\mu\nu\rho\sigma}
\left(\delta A_\mu{}^M {\cal D}_{\nu} {\cal F}_{\rho\sigma}{}_M
+
{\cal F}_{\mu\nu}{}_M
\Big(
 6 (t^\alpha)^{MN}\partial_N\Delta B_{\rho\sigma \alpha}
+\frac14\Omega^{MN}\Delta B_{\rho\sigma N}
\Big)\right)
\,, \qquad 
\eea
which takes a covariant form in terms of the general variations introduced in (\ref{DeltaB}).
From this expression it is straightforward to explicitly verify gauge invariance under 
$\Lambda$ and $\Xi$ transformations (\ref{gaugeL}), (\ref{tensor_gauge7}).

Variation of the combined Lagrangian ${\cal L}_{\rm YM}+{\cal L}_{\rm top}$ w.r.t.\ the two-forms
consistently reproduces parts of the duality equation (\ref{duality}).
More precisely, variation w.r.t.\ $B_{\mu\nu\,\alpha}$ yields the duality equation under internal derivatives 
$(t^\alpha)^{MN}\partial_N$ 
whereas variation w.r.t.\ $B_{\mu\nu\,M}$ formally seems to give all of (\ref{duality}), however one must
take into account that this field itself is constrained by (\ref{constraintsB}), such that the variation of its components
is not independent.

Concerning the Lagrangian of the gauge field sector, the sum ${\cal L}_{\rm YM}+{\cal L}_{\rm top}$
constitutes an incomplete (or `pseudo-')action that must be amended
by the additional first order duality equations (\ref{duality}). This is in the spirit of the `democratic formulation' 
of supergravities~\cite{Bergshoeff:2001pv}. In reality we are thus working on the level of the field equations
and simply introduce this Lagrangian as a convenient tool to verify symmetries of the field equations in a compact way. 
Alternatively, one may switch to a true Lagrangian formulation in the standard fashion~\cite{Cremmer:1979up,Gaillard:1981rj} 
by choosing a symplectic frame
that selects 28 electric vector fields $A_\mu{}^\Lambda$, breaking the matrix ${\cal M}_{MN}$ into
\bea
{\cal M}_{MN} &=& 
\left(
\begin{array}{cc}
{\cal M}_{\Lambda\Sigma} & {\cal M}_{\Lambda}{}^\Sigma\\
{\cal M}^{\Lambda}{}_{\Sigma} & {\cal M}^{\Lambda\Sigma}
\end{array}
\right)
~\equiv~
\left(
\begin{array}{cc}
({\cal I}+{\cal R}{\cal I}^{-1}{\cal R})_{\Lambda\Sigma} & -({\cal R}{\cal I}^{-1})_{\Lambda}{}^\Sigma\\
-({\cal I}^{-1}{\cal R})^{\Lambda}{}_{\Sigma} & ({\cal I}^{-1})^{\Lambda\Sigma}
\end{array}
\right)\;,
\label{LYMelec}
\eea
and replacing the kinetic term (\ref{LYM}) by
\bea
{\cal L}_{\rm YM}&=&-\frac{1}{4}\,e\, {\cal I}_{MN}\,{\cal F}^{\mu\nu M}{\cal F}_{\mu\nu}{}^{N}
-\frac{1}{8}\,\varepsilon^{\mu\nu\rho\sigma}\, {\cal R}_{MN}\,{\cal F}_{\mu\nu}{}^{M}{\cal F}_{\rho\sigma}{}^{N}
\;.
\label{LYM28}
\eea  
The topological term then is modified similar to the structure given in~\cite{deWit:2007mt} that treats asymmetrically
the electric and magnetic vector fields. The resulting Lagrangian carries 28 electric vectors with proper kinetic term (\ref{LYM28})
and 28 magnetic duals that only appear in covariant derivatives and the topological term. Its field equations are equivalent to those
we have been discussing above. For this paper, we prefer to work on the level of the field equations (or equivalently with
the `pseudo'-action (\ref{LYM})) since that formulation retains the manifest ${\rm E}_{7(7)}$ covariance.
  
Let us discuss the field equations of the vector/tensor system.  Taking the exterior derivative of (\ref{duality})
and using the Bianchi identity (\ref{Bianchi7}) one obtains second order field equations for the vector fields
\bea
{\cal D}_\nu \left( e\,{\cal M}_{MN}\,{\cal F}^{\mu\nu}{}^N \right)
&=&
-
2\,\varepsilon^{\mu\nu\rho\sigma} 
(t^\alpha)_M{}^{N}\partial_N {\cal H}_{\nu\rho\sigma\,\alpha}
+
\frac1{12}\,\varepsilon^{\mu\nu\rho\sigma} \,{\cal H}_{\nu\rho\sigma\,M}\;.
\label{derdual}
\eea
We may compare this equation to the field equations obtained from variation of
the Lagrangian (\ref{LYM}), (\ref{Ltop})
\bea
{\cal D}_\nu\,\left( e\,{\cal M}_{MN}\,{\cal F}^{\mu\nu}{}^N\right) &=& 
2e\,\left(\widehat{J}^\mu{}_{M} +{\cal J}^\mu{}_M \right)
-\frac12\,\varepsilon^{\mu\nu\rho\sigma}\,{\cal D}_\nu {\cal F}_{\rho\sigma\,M}
\label{eomV}
\eea
with the gravitational and matter currents defined by general variation w.r.t.\ the vector fields
\bea
\delta_{A} {\cal L}_{\rm EH} &\equiv& 
e\,\widehat{J}^\mu{}_{M}\,\delta A_\mu{}^M\;,\qquad
\delta_{A} {\cal L}_{\rm sc} ~\equiv~ e\,{\cal J}^\mu{}_{M}\,\delta A_\mu{}^M
\;,
\eea
e.g.~explicitly
\bea\label{STep}
 {\cal J}^\mu{}_{M} &=& 
 e^{-1}\,\partial_N\left(e\, {\cal D}^\mu {\cal M}^{NP} {\cal M}_{MP} \right)
 -\frac{1}{24}\,{\cal D}^\mu {\cal M}^{KL} \partial_M {\cal M}_{KL}
 \;.
\eea
Combining (\ref{derdual}) and (\ref{eomV}), 
we obtain the duality equations between scalar and tensor fields
\bea
e\,\widehat{J}^\mu{}_{M} +e\,{\cal J}^\mu{}_M 
&=&
-2\,\varepsilon^{\mu\nu\rho\sigma} 
(t^\alpha)_M{}^{N}\partial_N {\cal H}_{\nu\rho\sigma\,\alpha}
+
\frac1{12}\,\varepsilon^{\mu\nu\rho\sigma} \,{\cal H}_{\nu\rho\sigma\,M}
\;.
\eea
Inserting (\ref{STep}) we can project this equation onto its irreducible parts
and obtain 
\bea
e\,\widehat{J}^\mu{}_{M} -\frac{1}{24}\,e\,{\cal D}^\mu {\cal M}^{KL} \partial_M {\cal M}_{KL} 
&=&
\frac1{12}\,\varepsilon^{\mu\nu\rho\sigma} \,{\cal H}_{\nu\rho\sigma\,M}
\;,\nonumber\\
-\frac12\,(t_\alpha)_K{}^L\,\left(e\, {\cal D}^\mu {\cal M}^{KP} {\cal M}_{LP} \right)
&=&
\varepsilon^{\mu\nu\rho\sigma} 
 {\cal H}_{\nu\rho\sigma\,\alpha} 
 \;.
 \label{dualH}
\eea
More precisely, the second equation only arises under projection with the 
derivatives $(t^\alpha)^{MN}\partial_N$.

\paragraph{The potential}

Finally, we discuss the last term in the EFT action (\ref{finalaction}).  The potential $V$ is 
a function of the external metric $g_{\mu\nu}$ and the internal metric ${\cal M}_{MN}$ given by
\be\label{fullpotential}
 \begin{split}
  V \ = \ &-\frac{1}{48}{\cal M}^{MN}\partial_M{\cal M}^{KL}\,\partial_N{\cal M}_{KL}+\frac{1}{2} {\cal M}^{MN}\partial_M{\cal M}^{KL}\partial_L{\cal M}_{NK}\\
  &-\frac{1}{2}g^{-1}\partial_Mg\,\partial_N{\cal M}^{MN}-\frac{1}{4}  {\cal M}^{MN}g^{-1}\partial_Mg\,g^{-1}\partial_Ng
  -\frac{1}{4}{\cal M}^{MN}\partial_Mg^{\mu\nu}\partial_N g_{\mu\nu}\;. 
 \end{split} 
 \ee 
The relative coefficients in here are determined by $\Lambda^M$ gauge invariance, in
a computation that is analogous to the ${\rm E}_{6(6)}$ case presented in \cite{Hohm:2013vpa}
and that we briefly sketch in the following. We first note that acting with $\partial_M$ on an ${\rm E}_{7(7)}$ scalar $S$
adds a density weight of $-\frac{1}{2}$. Consider its variation
$\delta_{\Lambda}S=\Lambda^N\partial_NS$. It can then be easily checked by writing out 
the projector (\ref{adjproj}) that its partial derivative transforms covariantly as
 \be
  \delta_{\Lambda}(\partial_MS) \ = \ \mathbb{L}_{\Lambda}(\partial_MS)\;, \quad {\rm where} \qquad
  \lambda(\partial_MS) \ = \ -\frac{1}{2}\;, 
 \ee
i.e., as a co-vector density of weight $\lambda=-\frac{1}{2}$. Similarly, while ${\cal M}$ is a tensor 
of weight zero, its partial derivatives $\partial{\cal M}$ carry a weight of $-\frac{1}{2}$, which is 
precisely the right weight to combine with the weight 2 of the vierbein determinant $e$ to a total weight 
of $1$ for the potential term, as needed for gauge invariance of the action. In contrast to a scalar however, the partial derivative 
$\partial{\cal M}$ receives also various non-covariant terms whose cancellation needs to be 
verified explicitly. A direct computation gives for the first term in (\ref{fullpotential}), up to boundary terms,  
 \be\label{firstVar}
  \delta_{\Lambda}\Big(-\frac{1}{48}e{\cal M}^{MN}\partial_M{\cal M}^{KL}\,\partial_N{\cal M}_{KL}\Big)
  \ = \ e\partial_M\partial_R\Lambda^P\,{\cal M}^{MN}{\cal M}^{LR}\partial_N{\cal M}_{PL}\;.
 \ee 
For this computation one has to use that ${\cal M}^{-1}\partial{\cal M}$ takes values in the Lie algebra of E$_{7(7)}$ 
so that the adjoint projector acts as the identity, 
 \be
  \mathbb{P}^{R}{}_{S}{}^{K}{}_{Q}\,{\cal M}^{QL}\partial_N{\cal M}_{KL} \ = \  
  {\cal M}^{RL}\partial_N{\cal M}_{SL}\;.
 \ee
For the second term in (\ref{fullpotential}) one finds after a straightforward calculation 
  \be\label{straightforwardcomp}
  \begin{split}
  \delta_{\Lambda}\Big(\,\frac{1}{2}e {\cal M}^{MN}\partial_M{\cal M}^{KL}\partial_L{\cal M}_{NK}\,\Big) \ = \ 
  &-e\partial_M\partial_R\Lambda^P\,{\cal M}^{MN}{\cal M}^{LR}\partial_N{\cal M}_{PL}\\
  &+e\partial_M\partial_P\Lambda^L\partial_L{\cal M}^{MP}+e\partial_M\partial_P\Lambda^P\partial_L{\cal M}^{ML}\\[0.2cm]
  &-12e \partial_M\partial_R\Lambda^P (t_{\alpha})^{KR}(t^{\alpha})_{PQ}{\cal M}^{QL}{\cal M}^{MN}\partial_L{\cal M}_{NK}\\[0.2cm]
  &-\frac{1}{2}e\partial_M\partial_R\Lambda^P \Omega^{KR}\Omega_{PQ}{\cal M}^{QL}{\cal M}^{MN}\partial_L{\cal M}_{NK}
  \\[0.2cm]
  \  = \   &-e\partial_M\partial_R\Lambda^P\,{\cal M}^{MN}{\cal M}^{LR}\partial_N{\cal M}_{PL}\\[0.2cm]
  &+e\partial_M\partial_P\Lambda^L\partial_L{\cal M}^{MP}+e\partial_M\partial_P\Lambda^P\partial_L{\cal M}^{ML}
  \;. 
  \end{split}
  \ee
In the second equality we used again that the current $(J_L)^M{}_{K}\equiv {\cal M}^{MN}\partial_L{\cal M}_{NK}$ is Lie algebra 
valued, which implies that the terms in the third and fourth line are zero. 
In order to see this we note  
 \be
  2(J_L)^{(M}{}_{K}(t_{\alpha})^{R)K} \ = \ 2(J_{L})^{\beta}(t_{\beta})^{(M}{}_{K} (t_{\alpha})^{R)K}
  \ = \ (J_{L})^{\beta} f_{\beta\alpha}{}^{\gamma}(t_{\gamma})^{MR}\;, 
 \ee 
where we expanded the current into the basis $t_{\alpha}$ and used  the invariance of $(t_{\alpha})^{MN}$
in the final step. This is precisely the structure in the third line of (\ref{straightforwardcomp}), 
where this term is contracted with $\partial_M\partial_R\Lambda^P$ and hence zero by the section constraint. 
 Similarly, in the fourth line in (\ref{straightforwardcomp}) the symplectic form $\Omega^{KR}$ raises an 
 index on the current, whose free indices are then contracted with $\partial_M\partial_R\Lambda^P$, 
 giving zero by the section constraint. 
With the final result in (\ref{straightforwardcomp}) we see that 
the cubic term in ${\cal M}$ cancels the term in (\ref{firstVar}). It is straightforward to verify that the remaining 
two terms cancel against the variations coming from the second line in the potential (\ref{fullpotential}), 
up to total derivatives, 
thus proving full gauge invariance of the potential term.

For comparison of the full result with the truncations that have been given in the literature~\cite{Hillmann:2009ci,Berman:2011jh,Coimbra:2011ky},\footnote{See also \cite{Aldazabal:2013mya,Cederwall:2013naa} 
for the geometric interpretation of these terms.}
we finally note that after the truncation that sets $g_{\mu\nu}=e^{2\Delta}\eta_{\mu\nu}$, the potential term reduces to
\bea
{\cal L}_{\rm pot} ~=~-eV & = &
  e^{4\Delta}\,\Big(\frac{1}{48}{\cal M}^{MN}\partial_M{\cal M}^{KL}\,\partial_N{\cal M}_{KL}
  -\frac{1}{2} {\cal M}^{MN}\partial_M{\cal M}^{KL}\partial_L{\cal M}_{NK}\nonumber\\
  &&{}\qquad\quad
  +4\partial_M \Delta\,\partial_N{\cal M}^{MN}
  +12  {\cal M}^{MN}\partial_M \Delta\,\partial_N \Delta \Big)
  \;,
  \label{potentialtrunc}
\eea
and can be rewritten in terms of the rescaled matrix $\widehat{\cal M}_{MN} \equiv e^{\gamma \Delta} {\cal M}_{MN}$.
It is important to note that (\ref{potentialtrunc}) remains ${\rm E}_{7(7)}$ invariant only upon keeping $\Delta$ as
an independent degree of freedom.

\paragraph{External diffeomorphisms}

The various terms of the EFT action (\ref{finalaction}) have been determined
by invariance under generalized internal $\Lambda^M$ diffeomorphisms. 
In contrast, the relative coefficients between these terms are determined by
invariance of the full action (or equations of motion)
under the remaining gauge symmetries, which are a covariantized 
version of the external $(3+1)$-dimensional diffeomorphisms with parameters $\xi^{\mu}(x,Y)$. 
For $Y$-independent parameter, external diffeomorphism invariance is manifest. 
On the other hand, gauge invariance for general $\xi^{\mu}(x,Y)$ determines all
equations of motion with no free parameter left. 
The gauge variations of vielbein, scalars and the vector fields are given by 
 \bea
 \delta_\xi e_{\mu}{}^{a} &=& \xi^{\nu}{\cal D}_{\nu}e_{\mu}{}^{a}
 + {\cal D}_{\mu}\xi^{\nu} e_{\nu}{}^{a}\;, \nonumber\\
\delta_\xi {\cal M}_{MN} &=& \xi^\mu \,{\cal D}_\mu {\cal M}_{MN}\;,\nonumber\\
\delta_\xi A_{\mu}{}^M &=& \xi^\nu\,{\cal F}_{\nu\mu}{}^M + {\cal M}^{MN}\,g_{\mu\nu} \,\partial_N \xi^\nu
\;,
 \label{skewD}
\eea
i.e.\ take the form of covariantized diffeomorphisms together with an additional ${\cal M}$-dependent contribution in $\delta A$,
that has likewise appeared in~\cite{Hohm:2013jma,Hohm:2013vpa}.  
Invariance of (\ref{finalaction}) can be shown in close analogy to the calculation for the ${\rm E}_{6(6)}$ 
case of~\cite{Hohm:2013vpa}. Instead of repeating this discussion, let us spend a few words
on the particularities of the ${\rm E}_{7(7)}$ case, i.e.\ generalized diffeomorphism invariance 
of the first-order duality relations (\ref{duality}) and the transformation laws for the two-form tensors.
The latter fields transform as
\bea
\Delta_\xi B_{\mu\nu\,\alpha} &=& \xi^\rho\,{\cal H}_{\mu\nu\rho\,\alpha}\;,\nonumber\\
\Delta_\xi B_{\mu\nu\,M} &=& \xi^\rho\,{\cal H}_{\mu\nu\rho\,M}
+2e\,{\varepsilon}_{\mu\nu\rho\sigma} g^{\sigma\tau} {\cal D}^\rho\left(g_{\tau\lambda}\partial_M \xi^\lambda\right) 
\;,
\label{skewB}
\eea
in terms of the covariant variations (\ref{DeltaB}). In particular, the variation of the constrained compensating tensor gauge field $B_{\mu\nu\,M}$
carries an additional non-covariant term that is required for gauge invariance of the equations of motion. We note that 
a similar term has appeared in the transformation laws of the constrained compensating (vector) gauge fields in the
three-dimensional formulation~\cite{Hohm:2013jma}. Moreover, the structure of the transformation rule is manifestly
consistent with the constraints (\ref{constraintsB}) on this field.
From (\ref{skewD}), (\ref{skewB}), we find the transformation law of the field strengths
\bea
\delta_\xi {\cal F}_{\mu\nu}{}^M &=&
{\cal L}_\xi\,{\cal F}_{\mu\nu}{}^M + 2
\left(
{\cal D}_{[\mu}{\cal M}^{MN} g_{\nu]\rho} - 6 (t^\alpha)^{MN}\,{\cal H}_{\mu\nu\rho\,\alpha}
\right) \partial_N\xi^\rho
\nonumber\\
&&{}
+2\,{\cal M}^{MN}{\cal D}_{[\mu} \left( g_{\nu]\rho} \partial_N \xi^\rho\right)
-
e\,{\varepsilon}_{\mu\nu\lambda\sigma} g^{\sigma\tau} \Omega^{MN}\, {\cal D}^\lambda\left(g_{\tau\rho}\partial_N \xi^\rho\right)\;, 
\eea
where the first term describes the standard transformation under (covariantized) diffeomorphisms.
On-shell, upon using the duality equations (\ref{dualH}),
this transformation may be rewritten in the compact form
\bea
\delta_\xi {\cal F}_{\mu\nu}{}^M &=&
{\cal L}_\xi\,{\cal F}_{\mu\nu}{}^M + 
{\cal Z}_{\mu\nu}{}^M -
\frac12\,e {\varepsilon}_{\mu\nu\rho\sigma}\,\Omega^{MN}{\cal M}_{NK}\,{\cal Z}^{\rho\sigma\,K}
\;,\nonumber\\
&&{}\mbox{with}\quad
{\cal Z}_{\mu\nu}{}^M ~\equiv~2\,
{\cal D}_{[\mu}\left(
{\cal M}^{MN} g_{\nu]\rho}  \partial_N \xi^\rho\right)
\;.
\label{delXF}
\eea
From this expression it is evident that 
the non-covariant terms in the variation of ${\cal F}_{\mu\nu}{}^M$ drop out when 
calculating the variation of the duality equation (\ref{duality}):
\bea
\delta_\xi {\cal E}_{\mu\nu}{}^M &=& {\cal L}_\xi {\cal E}_{\mu\nu}{}^M
\;,
\eea
thus the duality equation is duality covariant. More precisely, as we used (\ref{dualH}), 
it follows that the first-order duality relations transform into each other. 
The discussion shows that the extra terms in the variation of
(\ref{skewB}) are crucial for this covariance. Moreover, the calculation requires the precise form (\ref{dualH})
of the duality equation between scalars and tensors and thereby fixes the corresponding relative coefficients
in the action (\ref{finalaction}). Eventually, external diffeomorphism invariance of the complete set of equations of motion
fixes all the coefficients in (\ref{finalaction}) and the equations of motion.

\section{Embedding $D=11$ and Type IIB Supergravity}

In the previous sections, we have constructed the unique set of ${\rm E}_{7(7)}$-covariant 
equations of motion for the fields (\ref{fieldcontent}), that is invariant under generalized internal and external diffeomorphisms.
It remains to explicitly embed $D=11$ supergravity.
Evaluating the above field equations with an explicit appropriate solution of the section constraints (\ref{sectioncondition}),
one may recover the full dynamics of $D=11$ supergravity after rearranging the eleven-dimensional fields according to a
$4+7$ Kaluza-Klein split of the coordinates, but retaining the full dependence on all eleven coordinates. 

The relevant solution of the section condition is related to the splitting of coordinates according to the
decomposition of the fundamental representation of  ${\rm E}_{7(7)}$ under its maximal ${\rm GL}(7)$ subgroup:
\bea
{\bf 56} &\longrightarrow&
{7}_{+3}+ { 21}'_{+1}+{21}_{-1}+ {7}'_{-3}\;,
\nonumber\\
\left\{ Y^M \right\} &\longrightarrow&
\left\{ y^m, y_{mn}, y^{mn}, y_m \right\}
\;. 
\label{dec56A}
\eea
Here subscripts refer to the ${\rm GL}(1)$ weight, 
indices $m, n, \dots$ label the vector representation of ${\rm GL}(7)$,
and the coordinates $y^{mn}=y^{[mn]}$, $y_{mn}=y_{[mn]}$ are antisymmetric in their indices.
The adjoint representation breaks according to
\bea
{\rm GL}(7)\,\subset \,{\rm E}_{7(7)}\,:\quad {\bf 133}  &\rightarrow&
{{7}'}_{\!+4}
+{ 35}_{+2}+
{1}_0+{48}_0+
{{ 35}'}_{-2}+
{7}_{-4}
\;. 
\label{decompE7A}
\eea
The ${\rm GL}(1)$ grading of these decompositions shows immediately that
\bea
(t_\alpha)^{m\,n} &=& 0
\;,
\eea
since there is no generator of charge $+6$ in the adjoint representation.
Consequently, the section constraints (\ref{sectioncondition})
are solved by truncating the coordinate dependence of all fields and gauge parameters to the coordinates in the ${7}_{+3}$:
\bea
\Phi(x^\mu, Y^M) &\longrightarrow&
\Phi(x^\mu, y^m)\;,\qquad
\mbox{i.e.}\quad
\partial^{mn} \rightarrow 0\;,\quad
\partial_{mn} \rightarrow 0\;,\quad 
\partial^m \rightarrow 0\;.
\label{solconA}
\eea
Accordingly, for the compensating gauge field constrained by (\ref{constraintsB}) 
we set all but the associated 7 components $B_{\mu\nu\,m}$ to zero
\bea
B_{\mu\nu}{}^{mn} \rightarrow 0\;,\quad
B_{\mu\nu\,mn} \rightarrow 0\;,\quad 
B_{\mu\nu}{}^m \rightarrow 0\;.
\label{solconA1}
\eea

The various fields of $D=11$ supergravity are recovered by splitting the vector fields $A_\mu{}^M$ and the two-forms
$B_{\mu\nu\,\alpha}$, $B_{\mu\nu\,M}$ according to (\ref{dec56A}), (\ref{decompE7A}),
and parametrizing the scalar matrix ${\cal M}_{MN}= ({\cal V}{\cal V}^T)_{MN}$ in terms of a group-valued
vielbein ${\cal V}$, defined in triangular gauge according to \cite{Cremmer:1997ct} as
\bea
{\cal V} &\equiv& {\rm exp} \left[\phi\, t_{(0)}\right]\,{\cal V}_7\;
{\rm exp}\left[c_{kmn}\,t_{(+2)}^{kmn}\right]\,{\rm exp}\left[\epsilon^{klmnpqr}  c_{klmnpq}\, t_{(+4)\,r}\right]
\;.
\label{V56}
\eea
Here, $t_{(0)}$ is the E$_{7(7)}$ generator associated to the GL(1) grading, 
${\cal V}_7$ denotes a general element of the ${\rm SL}(7)$ subgroup,
whereas the $t_{(+n)}$ refer to the E$_{7(7)}$ generators of positive grading in 
(\ref{decompE7A}). All generators are evaluated in the
fundamental ${\bf 56}$ representation (\ref{dec56A}).
Upon choosing an explicit representation of the generators $(t_\alpha)_M{}^N$ in terms of
${\rm SL}(7)$ invariant tensors, splitting of all tensors according to (\ref{dec56A}), (\ref{decompE7A}),
and explicitly imposing (\ref{solconA}), the above ${\rm E}_{7(7)}$ covariant field equations can be mapped
into those of $D=11$ supergravity.
This requires redefinitions of all the form fields originating from the $11$-dimensional 3-form and 6-form in the usual 
Kaluza-Klein manner, i.e., flattening the world indices with the elfbein and then `un-flattening' with the vierbein 
$e_{\mu}{}^{a}$, as well as subsequent further non-linear field redefinitions and appropriate dualization of
some field components. 
We have gone through this exercise in all detail in the ${\rm E}_{6(6)}$-covariant construction \cite{Hohm:2013vpa}
and reproduced the full and untruncated action of eleven-dimensional supergravity.
Here, we will restrict the discussion to illustrating the novel features of the E$_{7(7)}$ case.

The scalar fields $c_{mnk}=c_{[mnk]}$ and $c_{mnklpq}=c_{[mnklpq]}$ parametrizing the matrix ${\cal M}_{MN}$ according to (\ref{V56}) have 
obvious origin in the internal components of the $11$-dimensional 3-form and 6-form. Let us consider the 56 vector fields,
splitting according to (\ref{dec56A}) above into
\bea
\left\{ A_\mu{}^M \right\} &\longrightarrow&
\left\{ A_\mu{}^m, A_{\mu\,mn}, A_\mu{}^{mn}, A_{\mu\,m} 
\right\}
\;.
\label{splitAA}
\eea 
The first 7 vector fields $A_\mu{}^m$ correspond to the $D=11$ Kaluza-Klein vectors, whereas the $21+21$ components
$A_{\mu\,mn}$ and $A_\mu{}^{mn}$ are related to the corresponding components of the $11$-dimensional 3-form and 6-form,
respectively. The last 7 vector fields $A_{\mu\,m}$ have no direct appearance in $D=11$ supergravity, but capture some of the degrees
of freedom of its dual graviton. Let us consider their role in some more detail.
Evaluating a generic covariant derivative (\ref{covDV}), upon taking the above solution of the section constraint and 
using the split (\ref{splitAA}), shows that most of the vector fields
only appear under internal derivatives $\partial_m$; more precisely, out of the 56 vectors $A_\mu{}^M$, 
the full connection only carries the following combinations of gauge fields
\bea
{\cal D}\left(\left\{A^M\right\}\right)  &=& 
{\cal D}\left(\left\{A^m, \partial_{[k} A_{mn]}, \partial_k A{}^{km}\right\}\right)
\;.
\label{connA}
\eea
In particular, the 7 vectors $A_{\mu\,m}$ drop out from all covariant derivatives. Moreover, a quick counting
of the independent vector field components in this connection yields
\bea
A^m\;:\; 7\;,\qquad
\partial_{[k} A_{mn]} \;:\; 15\;,\qquad
\partial_k A^{km}\;:\; 6\;.
\label{countA}
\eea
E.g.\ the 21 components $A_{mn}$ enter the connection (\ref{connA}) in a way invariant under transformations
$A_{mn}\rightarrow A_{mn}+\partial_{[m} a_{n]}$ which can be used to set 6 of these components (say the $A_{m7}$)
to zero, etc.. This counting
shows that in total $7+15+6=28$ out of the 56 vector fields participate in the connections,
a counting that is also consistent with~\cite{Berman:2012vc}. This is in precise agreement with
the general structure of maximal gauged supergravities~\cite{deWit:2007mt}, in which at most 28 vector fields
participate in the gauging of some non-abelian symmetry.
We may perform an analogous counting of the number of two-form components from $B_{\mu\nu\,\alpha}$ 
that actually appear in the covariant field strengths (\ref{modF7}) and find
\bea
\partial_{[m} B_{n]}\;:\; 6\;,\qquad
\partial_k B{}^{kmn} \;:\; 15\;.
\eea
Together with the 7 components surviving in $B_{\mu\nu\,M}$ after imposing (\ref{solconA1})
this makes a total of 28 2-forms entering the covariant field strengths ${\cal F}_{\mu\nu}{}^M$
and thereby the twisted first-order self-duality equations (\ref{duality})
and the action (\ref{finalaction}).
Again, this counting is in precise agreement with
the general structure of maximal gauged supergravities~\cite{deWit:2007mt}: the existence of
non-abelian self-duality equations requires a compensating 2-form per vector field 
participating in the gauging.

In order to reproduce the field equations of $D=11$ supergravity, second order field equations for the vector
fields can be read off from (\ref{LYM28}), upon first decomposing the matrix ${\cal M}_{MN}$ obtained from (\ref{V56}) 
according to (\ref{LYMelec}), with a specific choice of symplectic frame. 
Alternatively, 21 of the first-order self-duality equations (\ref{duality}) can be mapped directly to the corresponding
components of the $D=11$ duality equations between 3-form and 6-form.
The seven remaining self-duality equations are those featuring the vector field $A_{\mu\,m}$ which has no origin
in the standard formulation of $D=11$ supergravity and rather corresponds to components of the $D=11$ dual graviton.
Only their derivatives (such that $A_{\mu\,m}$ drops from the equations) can be matched to the $D=11$ 
second order field equations. In the E$_{7(7)}$ covariant formulation, these equations exist as first-order duality
equations by virtue of the surviving components $B_{\mu\nu\,m}$ of the covariantly constrained fields
$B_{\mu\nu\,M}$ (\ref{solconA1}), that play the role of compensating tensor gauge fields.

Let us finally briefly discuss the embedding of IIB supergravity.
Just as for the ${\rm E}_{6(6)}$ EFT \cite{Hohm:2013pua,Hohm:2013vpa}, there is
another inequivalent solution to the section conditions (\ref{sectioncondition})
that describes the embedding of the full ten-dimensional IIB 
theory~\cite{Schwarz:1983wa,Howe:1983sra} into the ${\rm E}_{7(7)}$ 
EFT.\footnote{An analogous solution of the ${\rm SL}(5)$ covariant section condition, 
corresponding to some three-dimensional truncation of type IIB,  
was discussed recently in the truncation of the theory to its potential term \cite{Blair:2013gqa}.}
In this case, the relevant maximal subgroup of  ${\rm E}_{7(7)}$ is 
${\rm GL}(6)\times {\rm SL}(2)$, under which the fundamental and adjoint representation decompose according to
\bea
{\bf 56} &\rightarrow&
({6},1)_{+2}+(6',2)_{+1}+(20,1)_0 +(6,2)_{-1}+ (6',1)_{-2}
\;,\label{decompE7B}\\
{\bf 133}  &\rightarrow&
({{ 1}},{ 2})_{\!+3}+
({{ 15}'},{ 1})_{\!+2}+
({{ 15}},{ 2})_{\!+1}+
({{ 35}}+{ 1},{ 1})_{0}+
({{ 15}'},{ 2})_{\!-1}+
({{ 15}},{ 1})_{\!-2}+
({{ 1}},{ 2})_{\!-3}
\;,
\nonumber
\eea
with the subscript denoting the ${\rm GL}(1)$ charge. 
With the corresponding split of coordinates and vector fields\footnote{Indices $m, n = 1, \dots, 6$ and $a=1, 2$,
label the fundamental representations of ${\rm SL}(6)$ and ${\rm SL}(2)$, respectively. The
coordinates $y_{kmn}=y_{[kmn]}$ and vector fields $A_{\mu\,kmn}=A_{\mu\,[kmn]}$ are antisymmetric in all their internal indices.}
\bea
\left\{ Y^M \right\} &\rightarrow&
\left\{ y^m, y_{m\,a}, y_{kmn}, y^{m\,a}, y_m \right\}
\;,\nonumber\\
\left\{ A_\mu{}^M \right\} &\rightarrow&
\left\{ A_\mu{}^m, A_{\mu\,m\,a}, A_{\mu\,kmn}, A_\mu{}^{m\,a}, A_{\mu\,m} \right\}
\;,
\label{splitAB}
\eea
it follows as above, that the constraints (\ref{sectioncondition}) and (\ref{constraintsB})
are solved by restricting the coordinate dependence of all fields to the 6 coordinates $y^m$
(of highest ${\rm GL}(1)$ charge),
and setting all but the associated 6 components of $B_{\mu\nu\,M}$ to zero
\bea
&&
\partial^{m\,a} \rightarrow 0\;,\quad
\partial^{kmn} \rightarrow 0\;,\quad 
\partial_{m\,a} \rightarrow 0\;,\quad
\partial^{m} \rightarrow 0\;,
\nonumber\\
&&
B^{m\,a} \rightarrow 0\;,\quad
B^{kmn} \rightarrow 0\;,\quad 
B_{m\,a} \rightarrow 0\;,\quad
B^{m} \rightarrow 0\;.
\label{solconB}
\eea
The set of IIB fields and equations of motion is recovered 
upon choosing an explicit representation of the generators $(t_\alpha)_M{}^N$ in terms of
${\rm SL}(6)\times {\rm SL}(2)$ invariant tensors, splitting of all fields and tensors according to (\ref{decompE7B}),
and explicitly imposing (\ref{solconB}).
As above, this requires the standard Kaluza-Klein redefinitions together with additional non-linear redefinitions 
of all the form fields and appropriate dualization of some field components. 
The scalar matrix ${\cal M}_{MN}= ({\cal V}{\cal V}^T)_{MN}$ in this case is most conveniently parametrized
in terms of a group-valued vielbein ${\cal V}$, defined in triangular gauge as
\bea
{\cal V} &\equiv& {\rm exp} \left[\phi\, t_{(0)}\right]\,{\cal V}_6\,{\cal V}_2\;
{\rm exp}\left[c_{mn\,a}\,t_{(+1)}^{mn\,a}\right]\,
{\rm exp}\left[\epsilon^{klmnpq}\,c_{klmn}\,t_{(+2)}{\,}_{pq}\right]
{\rm exp}\left[c_{a}\,t_{(+3)}^{a}\right]
\;.
\label{V56B}
\eea
Here, $t_{(0)}$ is the E$_{7(7)}$ generator associated to the GL(1) grading, 
${\cal V}_6$ and ${\cal V}_2$ denote general elements of the ${\rm SL}(6)$ and ${\rm SL}(2)$ subgroups,
respectively, and the $t_{(+n)}$ refer to the E$_{7(7)}$ generators of positive grading in 
(\ref{decompE7B}). All generators are evaluated in the fundamental ${\bf 56}$ representation.
The scalar fields $c_{mn\,a}=c_{[mn]\,a}$ and $c_{a}$ in (\ref{V56B}) descend from
the internal components of the $10$-dimensional 2-form doublet and its dual 6-form doublet. 
In turn, $c_{klmn}$ has its origin in the internal components of the (self-dual) four-form.
From the 56 vector fields, split according to (\ref{splitAB}), the first 6 vector fields $A_\mu{}^m$ 
correspond to the $D=10$ Kaluza-Klein vectors, whereas the $44$ components
$A_{\mu\,ma}$, $A_{\mu\,kmn}$, and $A_\mu{}^{ma}$ are related to the corresponding 
components of the $10$-dimensional $p$-forms. Again, the last 6 vector fields $A_{\mu\,m}$ 
have no direct appearance in IIB supergravity, but capture some of the degrees
of freedom of its dual graviton. 
Evaluating a generic covariant derivative (\ref{covDV}), 
with (\ref{solconB}) and (\ref{splitAB}), shows that 
these 6 vectors drop out from all covariant derivatives. More precisely, 
the full connection only carries the following combinations of gauge fields
\bea
{\cal D}\left(\left\{A^M\right\}\right)  &=& 
{\cal D}\left(\left\{A{}^m,
\partial_{[m}A_{n]a},
\partial_{[k}\,A_{lmn]},
\partial_m A^{m\,a}\right\}\right)
\;.
\label{connB}
\eea
Counting
of the independent components similar to (\ref{countA})
\bea
A^m\;:\; 6\;,\qquad
\partial_{[m} A_{n]a} \;:\; 2\cdot5\;,\qquad
\partial_{[k} A_{lmn]} \;:\; 10\;,\qquad
\partial_m A^{m\,a}\;:\; 2\;,
\eea
shows that also for IIB there are precisely 28 out of the 56 vector field components 
which appear in the connections (\ref{connB}). Similarly, evaluation of the expressions (\ref{modF7})
shows that in this case from the 133 $B_{\mu\nu\,\alpha}$, 
only the combinations
\bea
\partial_{m} B_{a}\;:\; 2\;,\qquad
\partial_{[k} B_{mn]} \;:\; 10\;,\qquad
\partial_k B^{km\,a} \;:\; 2\cdot5\;,
\eea
appear in the covariant field strengths (\ref{modF7}) and the action (\ref{finalaction}).
Together with the 6 components surviving in $B_{\mu\nu\,M}$ after imposing (\ref{solconB})
this again makes a total of 28 2-forms entering the  
twisted first-order self-duality equations (\ref{duality}) as compensating tensor gauge fields.

The first- and second-order field equations of type IIB supergravity are obtained from (\ref{duality56})
and (\ref{finalaction}) with the above split of fields, constraints (\ref{solconB}), field redefinitions
and appropriate dualization.  Again, we note that
the six self-duality equations from (\ref{duality56}) featuring the vector fields $A_{\mu\,m}$ have no direct origin
in IIB, as these vector fields correspond to components of the $D=10$ dual graviton.
Only their derivatives (such that $A_{\mu\,m}$ drops out from the equations) can be matched to the standard 
IIB field equations.

\section{Summary and Outlook}\label{sec:5}
In this paper we have spelled out the details of the E$_{7(7)}$ exceptional field theory. 
The main conceptual novelty of this case as compared to E$_{6(6)}$ is that, 
from the eleven-dimensional perspective, the  $7$ Kaluza-Klein vectors are introduced together 
with their on-shell duals, satisfying an electric-magnetic twisted self-duality relation. 
These on-shell duals thus correspond,  again from the eleven-dimensional perspective, to components 
of the dual graviton. Despite the no-go theorems of \cite{Bekaert:2002uh,Bekaert:2004dz}, it is possible to 
consistently include those fields in a non-linear theory by virtue of the simultaneous inclusion 
of compensating (2-form) gauge fields. This naturally follows from the structure of the tensor hierarchy,
and also gives a duality-covariant form of the mechanism introduced in~\cite{Boulanger:2008nd}. 
A crucial aspect of this mechanism is that the compensating gauge field itself is covariantly 
constrained in that it needs to satisfy  E$_{7(7)}$ covariant constraints that are of the same structural 
form as the section constraints. 

Although a deeper conceptual understanding of these constrained fields is certainly desirable, 
we have seen in the above construction of a fully E$_{7(7)}$ covariant 
formulation that their presence appears unavoidable. Recall that the need for such constrained 2-forms 
was an immediate consequence of the algebraic structure of the E$_{7(7)}$ E-bracket Jacobiator.
Equivalently, these fields were found indispensable for the definition of a gauge covariant field strength (\ref{modF7})
for the vector fields. 
As we have discussed, this nicely fits into a more general pattern of the tensor hierarchy of exceptional field theories:
For the E$_{6(6)}$ theory of \cite{Hohm:2013vpa} the necessity of introducing additional constrained compensating fields
appears at the level of $3$-forms (which, however, do not appear explicitly in the action). 
Similarly, in E$_{8(8)}$ EFT the compensating gauge field appears among the vector fields and 
can be viewed as an E$_{8(8)}$ gauge potential, again subject to E$_{8(8)}$ covariant constraints as found for the 
Ehlers SL$(2,\mathbb{R})$ subgroup in \cite{Hohm:2013jma}. 
Its presence also cures the seeming obstacle of non-closure of the algebra of
generalized diffeomorphisms~\cite{Berman:2012vc}.\footnote{
The details for the  E$_{8(8)}$ EFT will be presented in a separate publication.}
It is intriguing to observe that 
this purely group-theoretical origin of the constrained compensator fields in the tensor hierarchy
precisely matches (and in fact enables) the appearance of components of 
the eleven-dimensional  dual graviton field among the physical fields
of the exceptional field theories.
E.g.\ we have seen in (\ref{solconA1}) that in the embedding of eleven-dimensional supergravity the constraint
(\ref{constraintsB}) implies that all but seven components of the compensating two-form are identically zero.
These non-vanishing components are precisely those that couple via (\ref{modF7}) to the field strengths
associated to the seven vector fields originating from the eleven-dimensional dual graviton.
One may speculate that eventually the constraints, both the section constraint involving the coordinates 
and the constraints on the compensator fields, may be relaxed so that in particular the 
physical significance of the dual graviton may become more transparent.

Eleven-dimensional and type IIB supergravity naturally embed into the E$_{7(7)}$ EFT, as discussed in 
sec.~4. We leave a more detailed description of this embedding at the level of the action or the 
field equations to future work. 
This should include the formulation of the fermionic sector and the supersymmetry 
transformations, which in turn will also clarify the relation to the reformulation of de Wit and Nicolai \cite{deWit:1986mz}. 
A natural question, among many, then is which gauged ${\cal N}=8$ supergravities can be embedded, 
via the E$_{7(7)}$ EFT, into eleven-dimensional supergravity and which may require the extended 
$56$ E$_{7(7)}$ coordinates in a non-trivial fashion (perhaps after a suitable relaxation of the constraints). 
We leave these and other questions for future work.

\section*{Acknowledgments}
The work of O.H. is supported by the 
U.S. Department of Energy (DoE) under the cooperative 
research agreement DE-FG02-05ER41360 and a DFG Heisenberg fellowship. 
We would like to thank Hadi and Mahdi Godazgar, Hermann Nicolai  and Barton Zwiebach 
for useful comments and discussions.

\section*{Appendix}

\begin{appendix}

\section{Algebraic relations}
\label{app:relations}
In this appendix we collect a few important E$_{7(7)}$ relations. 
First, contracting the adjoint indices of two generators, we have the 
following relation:
\bea
(t_\alpha)_M{}^K (t^\alpha)_N{}^L &=&
\frac1{24}\,\delta_M^K\delta_N^L
+\frac1{12}\,\delta_M^L\delta_N^K
+(t_\alpha)_{MN} (t^\alpha)^{KL}
-\frac1{24} \,\Omega_{MN} \Omega^{KL}
\;,
\label{A1}
\eea
for the projector onto the adjoint representation.
Contracting two of the fundamental indices, the relation (\ref{A1}) gives 
\bea
(t_\alpha)_M{}^K (t^\alpha)_K{}^N &=& \frac{19}8\,\delta_M^N
\;.
\eea
There are also various higher-order relations among the generators,
which we list as
\bea
0 &=& 9 (t^\alpha)_M{}^K (t^\beta)_{KN} 
(t_\alpha)^{(PQ} (t_\beta)^{RS)}
+2 \,(t^\alpha)_{[M}{}^{(P} (t_\alpha)^{QR} \,\delta_{N]}^{S)}
-
\frac18\,\Omega_{MN}\,(t^\alpha)^{(PQ} (t_\alpha)^{RS)}
\;, 
\nonumber\\
0&=&
(t^\alpha)_{NL} (t_\alpha)^{M(K}\,(t_\beta)^{Q)L}
+\frac1{12}\,(t_\beta)^{M(K}\,\delta_N^{Q)} - \frac1{24}\,(t_\beta)_N{}^{(K}\Omega^{Q)M}
\nonumber\\
&&{}+\frac1{24} (t_\beta)^{KQ}\,\delta_N^M
+\frac12\,(t^\alpha)_{NL} (t_\alpha)^{KQ}\,(t_\beta)^{ML}
-\frac12\,(t^\alpha)^{ML} (t_\alpha)^{KQ}\,(t_\beta)_{NL}
\;,
\label{E7rel5}
\eea
and their contraction
\bea
(t^\alpha)^{ML} (t_\alpha)^{NQ}\,(t_\beta)_{NL} &=& -\frac78\,   (t_\beta)^{MQ}
\;.
\eea

\end{appendix}



\begin{thebibliography}{10}

\bibitem{Hohm:2013pua}
O.~Hohm and H.~Samtleben, { Exceptional form of ${D}=11$ supergravity},  {
  Phys.Rev.Lett.} { 111} (2013) 231601,
[\href{http://xxx.lanl.gov/abs/1308.1673}{{\tt 1308.1673}}].

\bibitem{Hohm:2013vpa}
O.~Hohm and H.~Samtleben, { Exceptional field theory {I}: ${E}_{6(6)}$
  covariant form of {M}-theory and type {IIB}},
\href{http://xxx.lanl.gov/abs/1312.0614}{{\tt 1312.0614}}.

\bibitem{Siegel:1993th}
W.~Siegel, { {Superspace duality in low-energy superstrings}},  { Phys.Rev.} {
  D48} (1993) 2826--2837,
[\href{http://xxx.lanl.gov/abs/hep-th/9305073}{{\tt hep-th/9305073}}].

\bibitem{Hull:2009mi}
C.~Hull and B.~Zwiebach, { Double field theory},  { JHEP} { 0909} (2009) 099,
[\href{http://xxx.lanl.gov/abs/0904.4664}{{\tt 0904.4664}}].

\bibitem{Hull:2009zb}
C.~Hull and B.~Zwiebach, { The gauge algebra of double field theory and
  {C}ourant brackets},  { JHEP} { 0909} (2009) 090,
[\href{http://xxx.lanl.gov/abs/0908.1792}{{\tt 0908.1792}}].

\bibitem{Hohm:2010jy}
O.~Hohm, C.~Hull, and B.~Zwiebach, { {Background independent action for double
  field theory}},  { JHEP} { 1007} (2010) 016,
[\href{http://xxx.lanl.gov/abs/1003.5027}{{\tt 1003.5027}}].

\bibitem{Hohm:2010pp}
O.~Hohm, C.~Hull, and B.~Zwiebach, { {Generalized metric formulation of double
  field theory}},  { JHEP} { 1008} (2010) 008,
[\href{http://xxx.lanl.gov/abs/1006.4823}{{\tt 1006.4823}}].

\bibitem{Hohm:2010xe}
O.~Hohm and S.~K. Kwak, { Frame-like geometry of double field theory},  {
  J.Phys.} { A44} (2011) 085404,
[\href{http://xxx.lanl.gov/abs/1011.4101}{{\tt 1011.4101}}].

\bibitem{Hohm:2013bwa}
O.~Hohm, D.~L\"ust, and B.~Zwiebach, { {The Spacetime of Double Field Theory:
  Review, Remarks, and Outlook}},
\href{http://xxx.lanl.gov/abs/1309.2977}{{\tt 1309.2977}}.

\bibitem{Cremmer:1978km}
E.~Cremmer, B.~Julia, and J.~Scherk, { Supergravity theory in 11 dimensions},
  { Phys. Lett.} { B76} (1978)
409--412.

\bibitem{Cremmer:1979up}
E.~Cremmer and B.~Julia, { The ${SO}(8)$ supergravity},  { Nucl. Phys.} { B159}
  (1979)
141.

\bibitem{West:2003fc}
P.~C. West, { {$E_{11}$}, {$SL(32)$} and central charges},  { Phys.Lett.} {
  B575} (2003) 333--342,
[\href{http://xxx.lanl.gov/abs/hep-th/0307098}{{\tt hep-th/0307098}}].

\bibitem{deWit:1982ig}
B.~de~Wit and H.~Nicolai, { ${N}=8$ supergravity},  { Nucl. Phys.} { B208}
  (1982)
323.

\bibitem{deWit:2007mt}
B.~de~Wit, H.~Samtleben, and M.~Trigiante, { {The maximal ${D} = 4$
  supergravities}},  { JHEP} { 06} (2007) 049,
[\href{http://xxx.lanl.gov/abs/arXiv:0705.2101}{{\tt arXiv:0705.2101}}].

\bibitem{deWit:2005hv}
B.~de~Wit and H.~Samtleben, { Gauged maximal supergravities and hierarchies of
  nonabelian vector-tensor systems},  { Fortschr. Phys.} { 53} (2005) 442--449,
[\href{http://xxx.lanl.gov/abs/hep-th/0501243}{{\tt hep-th/0501243}}].

\bibitem{deWit:2008ta}
B.~de~Wit, H.~Nicolai, and H.~Samtleben, { Gauged supergravities, tensor
  hierarchies, and {M}-theory},  { JHEP} { 0802} (2008) 044,
[\href{http://xxx.lanl.gov/abs/arXiv:0801.1294}{{\tt arXiv:0801.1294}}].

\bibitem{Hohm:2013nja}
O.~Hohm and H.~Samtleben, { {Gauge theory of Kaluza-Klein and winding modes}},
  { Phys.Rev.} { D88} (2013) 085005,
[\href{http://xxx.lanl.gov/abs/1307.0039}{{\tt 1307.0039}}].

\bibitem{Curtright:1980yk}
T.~Curtright, { Generalized gauge fields},  { Phys.Lett.} { B165} (1985)
304.

\bibitem{Hull:2000zn}
C.~Hull, { Strongly coupled gravity and duality},  { Nucl.Phys.} { B583} (2000)
  237--259,
[\href{http://xxx.lanl.gov/abs/hep-th/0004195}{{\tt hep-th/0004195}}].

\bibitem{West:2001as}
P.~C. West, { {${{E}}_{11}$ and {M} theory}},  { Class. Quant. Grav.} { 18}
  (2001) 4443--4460,
[\href{http://xxx.lanl.gov/abs/hep-th/0104081}{{\tt hep-th/0104081}}].

\bibitem{Hull:2001iu}
C.~Hull, { Duality in gravity and higher spin gauge fields},  { JHEP} { 0109}
  (2001) 027,
[\href{http://xxx.lanl.gov/abs/hep-th/0107149}{{\tt hep-th/0107149}}].

\bibitem{Bekaert:2002uh}
X.~Bekaert, N.~Boulanger, and M.~Henneaux, { {Consistent deformations of dual
  formulations of linearized gravity: A no-go result}},  { Phys.Rev.} { D67}
  (2003) 044010,
[\href{http://xxx.lanl.gov/abs/hep-th/0210278}{{\tt hep-th/0210278}}].

\bibitem{Bekaert:2004dz}
X.~Bekaert, N.~Boulanger, and S.~Cnockaert, { {No self-interaction for two-column massless fields}},  
{ J.Math.Phys.} { 46}
  (2005) 012303,
[\href{http://xxx.lanl.gov/abs/hep-th/0407102}{{\tt hep-th/0407102}}].


\bibitem{Boulanger:2008nd}
N.~Boulanger and O.~Hohm, { {Non-linear parent action and dual gravity}},  {
  Phys.Rev.} { D78} (2008) 064027,
[\href{http://xxx.lanl.gov/abs/0806.2775}{{\tt 0806.2775}}].

\bibitem{Hohm:2013jma}
O.~Hohm and H.~Samtleben, { {U-duality covariant gravity}},  { JHEP} { 1309}
  (2013) 080,
[\href{http://xxx.lanl.gov/abs/1307.0509}{{\tt 1307.0509}}].

\bibitem{Coimbra:2011ky}
A.~Coimbra, C.~Strickland-Constable, and D.~Waldram, { {$E_{d(d)} \times
  \mathbb{R}^+$ generalised geometry, connections and M theory}},
\href{http://xxx.lanl.gov/abs/1112.3989}{{\tt 1112.3989}}.

\bibitem{Berman:2012vc}
D.~S. Berman, M.~Cederwall, A.~Kleinschmidt, and D.~C. Thompson, { {The gauge
  structure of generalised diffeomorphisms}},  { JHEP} { 1301} (2013) 064,
[\href{http://xxx.lanl.gov/abs/1208.5884}{{\tt 1208.5884}}].

\bibitem{Bergshoeff:2001pv}
E.~Bergshoeff, R.~Kallosh, T.~Ortin, D.~Roest, and A.~Van~Proeyen, { New
  formulations of ${D}=10$ supersymmetry and {D8-O8} domain walls},  { Class.
  Quant. Grav.} { 18} (2001) 3359--3382,
[\href{http://xxx.lanl.gov/abs/hep-th/0103233}{{\tt hep-th/0103233}}].

\bibitem{Gaillard:1981rj}
M.~K. Gaillard and B.~Zumino, { Duality rotations for interacting fields},  {
  Nucl. Phys.} { B193} (1981)
221.

\bibitem{Hillmann:2009ci}
C.~Hillmann, { Generalized ${E}_{7(7)}$ coset dynamics and {$D=11$}
  supergravity},  { JHEP} { 0903} (2009) 135,
[\href{http://xxx.lanl.gov/abs/0901.1581}{{\tt 0901.1581}}].

\bibitem{Berman:2011jh}
D.~S. Berman, H.~Godazgar, M.~J. Perry, and P.~West, { Duality invariant
  actions and generalised geometry},  { JHEP} { 1202} (2012) 108,
[\href{http://xxx.lanl.gov/abs/1111.0459}{{\tt 1111.0459}}].

\bibitem{Aldazabal:2013mya} 
  G.~Aldazabal, M.~Grana, D.~MarquŽs and J.~A.~Rosabal, 
  { Extended geometry and gauged maximal supergravity},
  { JHEP} { 1306} (2013) 046, 
  [\href{http://xxx.lanl.gov/abs/1302.5419}{{\tt 1302.5419}}].
  
\bibitem{Cederwall:2013naa} 
  M.~Cederwall, J.~Edlund and A.~Karlsson,
  { Exceptional geometry and tensor fields},
  { JHEP} { 1307} (2013) 028, 
    [\href{http://xxx.lanl.gov/abs/1302.6736}{{\tt 1302.6736}}].
  
\bibitem{Cremmer:1997ct}
E.~Cremmer, B.~Julia, H.~Lu, and C.~N. Pope, { {Dualisation of dualities. I}},
  { Nucl. Phys.} { B523} (1998) 73--144,
[\href{http://xxx.lanl.gov/abs/hep-th/9710119}{{\tt hep-th/9710119}}].

\bibitem{Schwarz:1983wa}
J.~H. Schwarz and P.~C. West, { {Symmetries and transformations of chiral
  ${N}=2$ ${D}=10$ supergravity}},  { Phys. Lett.} { B126} (1983)
301.

\bibitem{Howe:1983sra}
P.~S. Howe and P.~C. West, { {The complete ${N}=2$, ${D}=10$ supergravity}},  {
  Nucl. Phys.} { B238} (1984)
181.

\bibitem{Blair:2013gqa}
C.~D.~A. Blair, E.~Malek, and J.-H. Park, { {M}-theory and {F}-theory from a
  duality manifest action},
\href{http://xxx.lanl.gov/abs/1311.5109}{{\tt 1311.5109}}.

\bibitem{deWit:1986mz}
B.~de~Wit and H.~Nicolai, { $d = 11$ supergravity with local {$SU(8)$}
  invariance},  { Nucl.Phys.} { B274} (1986)
363.

\end{thebibliography}


\providecommand{\href}[2]{#2}\begingroup\raggedright\endgroup

\end{document}